\documentclass[
 nofootinbib,
showpacs,showkeys,
 bibnotes,
 amsmath,amssymb,
 longbibliography,
 lengthcheck,
]{revtex4-1}
\usepackage{slashed}
\usepackage{graphicx}% Include figure files
\usepackage{dcolumn}% Align table columns on decimal point
\usepackage{bm}% bold math
\usepackage{hyperref}% add hypertext capabilities
\usepackage{appendix}

\begin{document}

\title{Cooper Pairs' Magnetic Moment in MCFL Color Superconductivity }% Force line breaks with \\

\author{Bo Feng}
 \email{bfeng@utep.edu}%Lines break automatically or can be forced with \\
\author{Efrain J. Ferrer}
 \email{ejferrer@utep.edu}
\author{Vivian de la Incera}
 \email{vincera@utep.edu}
\affiliation{%
 Department of Physics, University of Texas at El Paso, 500 W. University Ave., El Paso, TX 79968, USA
}%



\date{\today}% It is always \today, today,

\begin{abstract}
We investigate the effect of the alignment of the magnetic moments of Cooper pairs of charged quarks that form at high density in three-flavor quark matter. The high density phase of this matter in the presence of a magnetic field is known to be the Magnetic Color-Flavor-Locked (MCFL) phase of color superconductivity. We derive the Fierz identities of the theory and show how the explicit breaking of the rotational symmetry by the uniform magnetic field opens new channels of interactions and allows the formation of a new diquark condensate. The new order parameter is a spin-1 diquark condensate proportional to the component in the field direction of the average magnetic moment of the pairs of charged quarks. In the region of large fields, the new condensate's magnitude becomes comparable to the larger of the two scalar gaps. Since there is no solution of the gap equations with nonzero scalar gaps and zero value of this magnetic moment condensate, its presence in the MCFL phase is unavoidable. This is consistent with the fact that the extra condensate does not break any symmetry that has not already been broken by the known MCFL gaps. The spin-1 condensate enhances the condensation energy of pairs formed by charged quarks and the magnetization of the system. We discuss the possible consequences of the new order parameter on the issue of the chromomagnetic instability that appears in color superconductivity at moderate density.

%\begin{description}
%\end{description}
\end{abstract}

%Use showkeys class option if keyword
\maketitle

%\tableofcontents

\section{Introduction}
It has been suggested long ago that a phase transition to color superconducting (CS) quark matter may take place at
high baryon density and sufficiently low temperature \cite{ASRS}. In nature, the most likely place to
accomplish CS is the core of compact stars. The typical density inside a compact star can be of the order of several times the nuclear density. In such media,
quarks may be released from hadrons and thanks to the low star's temperature caused by the
rapid cooling through the neutrino emission taking place after the supernova collapse, they form CS Cooper pairs.

Compact stars, on the other hand, are very magnetized objects. From the measured periods and spin down of soft-gamma repeaters (SGR) and anomalous X-ray pulsars (AXP), as well as the observed X-ray luminosities of AXP, it has been found that a certain class of neutron stars named magnetars can have surface magnetic fields as large as $10^{14}-10^{16}$ G \cite{Magnetars}. Moreover, since the stellar medium has a very high electric conductivity, the magnetic flux should be conserved. Hence,
it is expected an increase of the magnetic field strength with increasing matter density, and consequently a much stronger magnetic field in the stars'
core. Nevertheless, the interior magnetic fields of neutron stars are not directly accessible to observation, so one can only estimate their values with heuristic methods. Estimates based on macroscopic and microscopic analysis, for nuclear \cite{virial}, and quark matter considering both gravitationally bound and self-bound stars \cite{EoS-H}, have led to maximum fields within the range $10^{18}-10^{20}$ G, depending if the inner medium is formed by neutrons \cite{virial}, or quarks \cite{EoS-H}.

The presence of a strong field in the star core can in principle modify the properties of the matter phase there and lead to observable signatures. Therefore, the investigation of the properties of very dense matter in the presence of strong magnetic fields is of interest not just from a fundamental point of view, but it is also closely connected to the physics of strongly magnetized neutron stars.

 %Thus, it has been estimated by direct application of the virial theorem for gravitational bound star with constant mass density and uniform magnetic fields that the
%inner fields of compact stars with masses $M\sim 1.4 M_\odot$ and radius $R\sim 10^{-4} R_\odot$ are as large as $\sim 10^{18}$ $G$ \cite{virial}. Relaxing the previous
%conditions, it has been shown recently that the estimated inner-field limit can be even higher ($10^{20}$ G) \cite{EoS-H} for self-bound stars, as well as for %gravitational-bound ones not subjected to uniform distributions of field and matter.

The effect of a strong magnetic field in the phase structure and gap magnitude of a color superconductor was first studied in \cite{mCFL}. There, it was shown that the presence of a magnetic field changes the color-flavor-locking (CFL) phase that realizes at large densities, producing a difference between the gap that gets contributions from pairs of oppositely charged quarks, denoted in \cite{mCFL} by $\Delta_{B}$, and the one that only gets contributions from pairs of neutral quarks, denoted by $\Delta$. Charged and neutral in this context refers to the electric charge associated to the "rotated" or in-medium electromagnetic group that remains unbroken in the CFL phase \cite{CFL}. In this phase, even though the original electromagnetic $U(1)_{em}$ symmetry is broken by the pairing, there is a residual symmetry that gives rise to a new long-range field composed by a linear combination of the conventional photon field and the 8th gluon field \cite{CFL,ABR}. The CFL pairs remain neutral with respect to this new "rotated" electromagnetism, and hence, the color superconductor can be penetrated by a "rotated" magnetic field without the restriction of the Meissner effect \cite{Dirk}. The mixing angle between the original electromagnetic field and the 8th gluon is such that a regular magnetic field penetrates the superconductor through its rotated magnetic field component with almost unchanged strength. The new phase that forms in the presence of the magnetic field also has color-flavor-locking, but with a smaller symmetry group $SU(2)_{C+L+R}$ and this change is reflected in the splitting of the gaps. This magnetic phase was called Magnetic CFL (MCFL) \cite{mCFL}. Similar to the CFL case, the MCFL phase is invariant under the rotated electromagnetic group.

At sufficiently strong magnetic fields (${\tilde e}{\tilde B}\gtrsim \mu^2$), the separation between the two gaps of the MCFL phase becomes significant, as the energy gap with contributions from pairs formed by rotated-charged quarks raises with the field while the other gap decreases with it \cite{mCFL}. However, contrary to what our na\"{\i}ve intuition might indicate, a magnetic field does not need to be of the order of the baryon chemical potential to produce a noticeable effect in a color superconductor. As shown in \cite{Phases}, the color superconductor is characterized by various scales and different physics can occur at field strengths comparable to each of these scales. Specifically, the MCFL phase is entirely attained when the field strength induces masses equal or larger than those of the charged Goldstones' constituents (i.e. twice of the CFL gap), and hence allowing the decay of this bound states. This effect is manifested at moderate magnetic fields (henceforth magnetic field always means rotated magnetic field), through the fact that all the energy gaps exhibit oscillations with respect to ${\tilde e}{\tilde B}/\mu^2<1$ \cite{NS, FW}. This behavior is due to the well known de Haas-van Alphen effect produced by the discreteness of the Landau levels in an external magnetic field, which becomes noticeable at field values beyond the threshold field needed to induce the Goldstone-mesons decay.

Although there are several interesting alternatives for CS pairing patterns possessing nontrivial magnetic properties as well \cite{BF}, in this paper we restrict ourselves to the color antitriplet channel that is known to be favored for three massless quark flavors at high densities. In the presence of a magnetic field the ground state of this system is in the MCFL phase \cite{mCFL}. However, it should be noticed that in this phase the Cooper pairs formed by charged quarks have nonzero magnetic moment because the quarks in the pair not only have opposite charge but also opposite spin. It is reasonable to expect that the MCFL ground state could have a net nonzero magnetic moment $\widetilde{\mu}$ which could couple with the applied magnetic field as $\sim \widetilde{\mu}\cdot \widetilde{B}$. This extra magnetic energy contribution of the diquark ground state would be reflected in the existence of a new order parameter $\Delta_M$ in addition to $\Delta$ and $\Delta_{B}$. The main goal of the present paper is to explore precisely this possibility.

Symmetry arguments can give additional insight on the likelihood of an extra gap $\Delta_M$ in the MCFL phase. The presence of a magnetic field breaks the spatial rotational symmetry $O(3)$ to the subgroup of rotations $O(2)$ about the axis parallel to the field. As will be shown in this work, this fact opens new attractive pairing channels that are not available in the CFL phase. One of these channels has symmetric Dirac structure $C\gamma_{5}\gamma_{1}\gamma_{2}$. As it will be discussed in detail later, a condensate with this structure is a spin-1 condensate with zero spin projection in the direction of the magnetic field. A diquark condensate with this new structure would have the same energy dimension as the regular gaps and would account for a magnetic energy associated to the projection of the magnetic moment of the ground state along the external field. With this understanding, we will call it the magnetic-moment condensate, because it should be proportional to the magnetic moment of the ground state. We will see that this interpretation is in agreement with the way $\Delta_M$ enters in the dispersion relations of the charged quasiparticles. As it will be shown, the color-flavor structure of $\Delta_M$ can be chosen in such a way that this condensate does not break any symmetry that has not already been broken by the gaps $\Delta$ and $\Delta_{B}$. Hence, such a magnetic-moment condensate is not symmetry-protected and in principle can be  different from zero. This resembles the reasons that led to the existence of the symmetric gaps in the CFL phase. The difference is that the color channel for the symmetric gaps is repulsive, so their magnitudes are typically much smaller than that of the antisymmetric ones. This is not the case with the magnetic moment, because the color channel here is going to be the same antitriplet attractive channel that favors the antisymmetric gaps. Nevertheless, since the magnetic-moment condensate is a direct consequence of the external magnetic field, we will see that its magnitude becomes comparable to the energy gaps only at strong field values.

Despite some fundamental differences, the MCFL scenario described above shares a few similarities with the dynamical generation of an anomalous magnetic moment recently found in massless QED \cite{magmoment}. Akin to the Cooper pairs of oppositely charged quarks in the MCFL phase, the fermion and antifermion that pair in massless QED also have opposite charges and spins and hence carries a net magnetic moment. A dynamical magnetic moment term in the QED Lagrangian does not break any symmetry that has not already been broken by the chiral condensate. Therefore, once the chiral condensate is formed due to the magnetic catalysis of chiral symmetry breaking \cite{catalysis, Miransky, Leung}, the simultaneous formation of a dynamical mass and a dynamical magnetic moment is unavoidable \cite{magmoment}. As we shall show below, the simultaneous generation of the MCFL gaps and the magnetic-moment condensate $\Delta_M$ is also unavoidable in the MCFL phase of color superconductivity. There is simply no consistent solution with nonzero energy gaps and zero $\Delta_M$.

More important, even though the solution for $\Delta_M$ is negligible at small magnetic fields, it becomes comparable to $\Delta_{B}$ at strong fields, where the LLL contribution is more relevant, indicating that the main contribution to $\Delta_M$ comes from the pairs formed by quarks that lie in the LLL. In addition, the gap $\Delta_{B}$ is actually larger (about twice at very strong fields) when $\Delta_M$ is taken into consideration in the analysis than when it is ignored. These two facts make the new parameter $\Delta_M$ of particular relevance for the potential realization of the MCFL phase in the core of magnetars. As known, the CFL phase may become chromomagnetic unstable at intermediate densities due to the pairing stress generated by the strange quark mass and the color and electric neutrality conditions \cite{{Fuku}}. In this context, any effect that can augment the effective gap magnitude will contribute to stabilize the phase by pushing the emergence of the instability to smaller regions of densities. Since magnetars have the strongest surface fields, they should also have the strongest fields in the core, so they are the best candidates for the realization of the magnetic CFL phase studied in this paper. The larger value of the gap $\Delta_{B}$ at strong fields will be also reflected in a larger critical temperature for this color superconducting phase, a property that opens yet another possibility for the realization of the MCFL state because the conditions of high densities, low temperatures, and strong magnetic fields will likely coexist in the planned low-temperature/high-density experiments at NICA@JNIR, CBM@FAIR and low-energy RIHC \cite{futureexp}.

In this paper, we investigate the generation of a magnetic moment condensate in the MCFL phase of color superconductivity, that is, in a three-flavor quark system at high density and in the presence of an external magnetic field.  We perform our calculations within a Nambu-Jona-Lasinio (NJL) model in the presence of an external magnetic field. The outline of the paper is as follows. In Sec. II, we discuss the new pairing channel that becomes available in the MCFL phase due to the external magnetic field, allowing for a new parameter (the magnetic-moment condensate $\Delta_M$)  in the gap matrix. This new gap is a spin-1 condensate with zero spin projection ($M_S=0$) along the field. To explore the simultaneous solution for $\Delta_M$, $\Delta$ and $\Delta_{B}$, we introduce a new MCFL ansatz for the color-flavor-Dirac structure of the gap matrix that contains the two usual energy gaps and the magnetic-moment condensate. This matrix respects all the symmetries of the old MCFL phase \cite{mCFL}, i.e., the locking between color and flavor described by the group $SU(2)_{C+L+R}$ and the invariance under the rotations in the plane perpendicular to the external field.  We then obtain the corresponding free energy and the minimum equations for the three order parameters in Sec. III. In Sec IV, we present our main numerical solutions for all the gaps parameters. A discussion of the results and our main conclusions are given in Sec. IV. In Appendix A, the details of the Fierz transformation under both Lorentz and rotational symmetry breaking is given. In Appendix B, the spin-1 nature of the magnetic-moment condensate is discussed. The Lagrangian density for lowest Landau levels is obtained in Appendix C using a chiral-spin representation.

\section{Gap Structure in MCFL with Magnetic Moment}

If we neglect the masses of all three light quarks at ultra-high densities, the general CFL pairing ansatz \cite{CFL} spontaneously breaks the full QCD symmetry group $SU(3)_C\times SU(3)_L\times SU(3)_R\times U(1)_B$ into the diagonal $SU(3)_{C+L+R}$ subgroup. The original electromagnetism $U(1)_Q$ is an implicit subgroup of flavor symmetry. Within the residual gauge group $SU(3)_{C+L+R}$, there exists a $U(1)_{\tilde Q}$ transformation, $\psi\rightarrow e^{-i\tilde Q\theta}\psi$ ($\psi$ is the quark field) that leaves the pairing invariant. The generator of this remanent symmetry is
\begin{equation}
\tilde Q=Q\times 1+1\times\frac{1}{\sqrt{3}}T_8.
\label{generator}
\end{equation}
with Q the conventional electromagnetic charge of quarks and $T_8$ the 8th Gell-Mann matrix. Upon the choice of the representation of the matrices $Q={\rm diag}(-1/3,-1/3,2/3)$ for $(s, d, u)$ flavors and $T_8={\rm diag}(-1/\sqrt{3},-1/\sqrt{3}, 2/\sqrt{3})$ for $(b, g, r)$ colors, we find that the $\tilde Q$ charges of the different quarks are \cite{mCFL}
\vspace{\baselineskip}
\begin{ruledtabular}
\begin{tabular}{ccccccccc}
\textrm{$s_b$}&
\textrm{$s_g$}&
\textrm{$s_r$}&
\textrm{$d_b$}&
\textrm{$d_g$}&
\textrm{$d_r$}&
\textrm{$u_b$}&
\textrm{$u_g$}&
\textrm{$u_r$}\\
\colrule
0 & 0 & $-$ & 0 & 0 & $-$ & $+$ & $+$ & 0\\
\end{tabular}
\end{ruledtabular}
\vspace{\baselineskip}

The eigenfield corresponding to the symmetry generated by (\ref{generator}) is given by a linear combination of the photon field $A_{\mu}$ and the 8th gluon field $G_{\mu}^8$ \cite{CFL, ABR}
\begin{equation}
\tilde A_{\mu}=\cos\theta A_{\mu}-\sin\theta G_{\mu}^8,
\label{rotated}
\end{equation}
The long-range field $\tilde A_{\mu}$ behaves in the CS medium as the new effective electromagnetic field. The corresponding orthogonal linear combination
\begin{equation}
\tilde G_{\mu}^8=\sin\theta A_{\mu}+\cos\theta G_{\mu}^8.
\end{equation}
will be massive. The mixing angle is defined by $\cos\theta=g/\sqrt{e^2/3+g^2}$, which is sufficiently small ($\theta\sim 1/40$) at moderate density ($g^2/(4\pi)\sim 1$). Effectively, the "rotated" photon (\ref{rotated}) mostly consists of the usual photon with a small admixture of the 8th gluon.

Clearly, the presence of an external magnetic field explicitly breaks the flavor symmetry from $SU(3)_L\times SU(3)_R$ to $SU(2)_L\times SU(2)_R$, since the different electric charge of the $s$ and $d$ quarks from the $u$ quark's, is manifested in the coupling with the external field. Assuming that the most energetically favored condensate is the one that preserves the highest degree of symmetry, it was proposed in \cite{mCFL} a condensate ansatz that in the presence of an external magnetic field spontaneously broke the symmetry to the color-flavor-locked subgroup $SU(2)_{C+L+R}$. This implies that the CFL gap parameters $\Delta_1$, $\Delta_2$ and $\Delta_3$, which describe $d-s$, $u-s$ and $u-d$ diquark pairs respectively, and which satisfy $\Delta_1=\Delta_2=\Delta_3$ in the CFL phase, change to
$\Delta=\Delta_1$ and $\Delta_B=\Delta_2=\Delta_3$ in the MCFL case. This is the MCFL ansatz \cite{mCFL}. Notice that even though all diquark pairs are neutral with respect to the rotated electromagnetic charge, they can be formed either by a pair of neutral or by a pair of opposite rotated charged quarks. Explicitly, the gap $\Delta$ gets only contribution from pairs of neutral quarks, while $\Delta_B$ has contributions from both charged and neutral quarks and should directly feel the background field through the minimal coupling of the quarks in the pair with $\tilde B$. Although the pairs contributing to $\Delta$ are all neutral, this gap is indirectly affected by the external field through the coupled gap equations. The symmetric gaps are also separated in two in the MCFL phase, so there is one $\Delta_{S}$ and one $\Delta_{SB}$ that were indeed considered in the derivations of Ref. \cite{mCFL}. In the  present work, however, we shall ignore them, as they are always smaller than their corresponding antisymmetric counterparts.

As previously mentioned, we expect a nonzero magnetic moment condensate to be present along with the MCFL energy gaps. That condensate can only exist if a proper channel opens up due to the external field. An external magnetic field (assumed here to be along the z-direction) introduces a normalized tensor $\widehat{F}^{\mu\nu}=\widehat{F}^{\mu\nu}/|B|$, with $\mu, \nu=1,2$. Because of this extra tensor in the theory, the metric tensor can be separated into transverse
\begin{equation}
g^{\mu\nu}_{\perp}= \widehat{F}^{\mu\rho}\widehat{F}_{\rho}^{\nu}
\end{equation}
and longitudinal
\begin{equation}
g^{\mu\nu}_{\parallel}= g^{\mu\nu}-g^{\mu\nu}_{\perp}
\end{equation}
components. Next, the finite density introduces yet another normalized vector in the theory, the four-velocity $u_{\mu}$, which in the rest frame reduces to $u_{\mu}=(1,0,0,0)$. With these structures, the four-fermion interaction term in the system Lagrangian density
\begin{equation}
{{\cal L}_{int}}=-G(\bar\psi\Gamma^a_\mu\psi)(\bar\psi\Gamma_a^\mu\psi),
\label{int-lagrangian}
\end{equation}
with quark-gluon vertex $\Gamma^a_\mu=\gamma_\mu \lambda^a$ (where $\lambda^a$'s are the generators of $SU(3)_C$ group), has Dirac contraction given by
\begin{equation}
\gamma^\mu \gamma_\mu = [au_\mu u_\nu+bg_{\mu\nu}^\bot +c(g_{\mu\nu}^\|-u_\mu u_\nu)]\gamma^\mu \gamma^\nu.
\label{contraction}
\end{equation}
Hence, one can write the four-fermion interaction (\ref{int-lagrangian}), in the rest frame, as three distinct terms
\begin{eqnarray}\label{4fermion-interact}
\nonumber {\cal L_{\textit{int}}}=&-&g_E({\bar\psi}\gamma_0\lambda^a\psi)({\bar\psi}\gamma_0\lambda^a\psi)
-g_M^\perp({\bar\psi}\gamma^\perp\lambda^a\psi)({\bar\psi}\gamma_\perp\lambda^a\psi)\\
&-&g_M^3({\bar\psi}\gamma^3\lambda^a\psi)({\bar\psi}\gamma_3\lambda^a\psi).
\end{eqnarray}
Using a Fierz transformation (see Appendix A for details) one can readily verify that the breaking of the Lorentz and rotational symmetries gives rise to new particle-particle channels of interaction, and in particular to $(\sigma^{ab}\gamma^5{\cal C})({\cal C}\gamma^5\sigma_{ab})$ (with ${\cal C}=i\gamma^2\gamma^0$ and $\sigma^{ab}=\frac{1}{2}[\gamma^a,\gamma^b]$, where $a,b=1,2$). Notice that the structure $\gamma^5\sigma^{ab}{\cal C}$ preserves Parity.
%Given that $\Sigma^3=\sigma^{12}$ is the spin operator along the z-axis, a condensate  $T=\langle \psi^T {\cal{C}} \Sigma^3 \gamma^5 \psi \rangle$ can be interpreted as the projection of a magnetic moment in the field direction.

\textit{What color-flavor structure can have the new condensate $\Delta_M=\langle \psi^T{\cal C} \gamma^5 \sigma_{12} \psi\rangle$ if it has to preserve the color-flavor symmetry of the known MCFL ground state?}
First, because we want to guarantee the strongest attractive channel, we choose it to be antisymmetric in color. Second, to ensure the total antisymmetry required by Pauli principle, and given the symmetric nature of ${\cal C}\gamma_5 \sigma_{ab}$ under transposition in Dirac indexes, $\Delta_M$ should be symmetric in flavor. Finally, it is natural to expect that this condensate will receive contributions from the same sector of Cooper pairs that contributes to the gap $\Delta_{B}$, that is, the color-flavor sector that contains pairs of charged quarks that minimally couple to the field in the Lagrangian. On the basis of all these considerations, we propose that the gap structure in the presence of a magnetic field takes the form
\begin{widetext}
\begin{equation}
\Phi^{\alpha\beta}_{ij}={\hat\Delta} \epsilon^{\alpha\beta 3}\epsilon_{ij3}+{\hat\Delta_B}(\epsilon^{\alpha\beta 1}\epsilon_{ij1}+\epsilon^{\alpha\beta 2}\epsilon_{ij2})
+{\hat\Delta_M}[\epsilon^{\alpha\beta 1}(\delta_{i2}\delta_{j3}+\delta_{i3}\delta_{j2})+\epsilon^{\alpha\beta 2}(\delta_{i1}\delta_{j3}+\delta_{i3}\delta_{j1})],
\label{generalansatz}
\end{equation}
\end{widetext}
where $\alpha,\beta$ and $i,j$ denote color and flavor indices respectively. Notice that all the coefficients in (\ref{generalansatz}) are actually matrices in Dirac space, for which we defined ${\hat\Delta}={\Delta{\cal C}\gamma_5}$, ${\hat\Delta_B}={\Delta_B{\cal C}\gamma_5}$ and ${\hat\Delta_M}={\Delta_M{\cal C}\gamma_5\sigma_{12}}$.

%Notice that all the coefficients in (\ref{generalansatz}) are actually implicit matrices in Dirac space, with structures ${\cal C}%\gamma^5\sigma_{12}$ for ${\cal T}$ and ${\cal C}\gamma^5$ for $\Delta$ and $\Delta_{B}$ respectively.

One can see from (\ref{generalansatz}) that the color-flavor structure of the gap $\Delta_M$ is different from that of the conventional gaps $\Delta$ and $\Delta_B$, which are antisymmetric in both color and flavor. It is easy to check that the ansatz (\ref{generalansatz}) is invariant under the same color-flavor symmetry than the MCFL ansatz \cite{mCFL}. That is, the $SU(2)_{C+L+R}$ symmetry, which requires the invariance of  $\Phi^{\alpha\beta}_{ij}$ under simultaneous flavor ($1\leftrightarrow 2$) and color ($1\leftrightarrow 2$) exchanges.

The symmetric Dirac structure of $\Delta_M$ corresponds to a spin-1 pairing of two quarks $\langle \psi^T{\cal C}\gamma^5\sigma_{12}\psi\rangle$. As shown in Appendix B, the specific spin-1 condensate we are considering has zero-spin projection ($M_S=0$) along the field direction. Thus, it corresponds to a symmetric wave function associated to pairs formed by quarks with opposite charges and opposite spins in the z-direction, and consequently with net magnetic moment projection in the $z$-direction. An order parameter with this same Dirac structure was considered in a 2SC model with no magnetic field in Ref. \cite{Buballa-M, Buballa}. In contrast to our case, where the rotational $O(3)$ symmetry has been already broken by the magnetic field, the  spin-1 condensate in \cite{Buballa-M} spontaneously breaks $O(3)$. Moreover, since only quarks of a single color participate in the condensate of \cite{Buballa-M}, it is a color-symmetric condensate which in addition breaks the remnant $\widetilde{U}(1)$ electromagnetic symmetry because has nonzero net rotated charge. In our case, as all the pairs remain neutral, the $\widetilde{U}(1)$ is not broken and there is no Meissner effect for the rotated electromagnetism.

 We have already argued in the Introduction why a new order parameter $\Delta_M$ is unavoidable on the basis of symmetry arguments, because the O(3) rotational symmetry is explicitly broken by the magnetic field to the O(2) group of rotations about the axis parallel to the field. The magnetic field dependence of this condensate also separates it from other spin-1 condensates found in the literature on color superconductivity \cite{Buballa-M}-\cite{Spin-1}. In those cases, the spin-1 condensate is always several orders of magnitude smaller than the spin-0 condensate. In our case, as we will show below, at sufficiently high magnetic fields the spin-1 condensate $\Delta_M$ is comparable in magnitude to the gap $\Delta_{B}$.

 Finally, because the $\Delta_M$ condensate does not break any other symmetry that was not already broken by the magnetic field and by the condensates $\Delta$ and $\Delta_B$, one has that the number of Goldstone fields of the MCFL phase \cite{mCFL, Phases} are not changed by the existence of this new order parameter.

\section{Free energy and gap equations}

In order to obtain the free energy and to derive the gap equations for the three unknown order parameters $\Delta$, $\Delta_B$ and $\Delta_M$, we use a massless three-flavor model with the local NJL-type interaction (\ref{int-lagrangian}). In this model, the interaction between quarks has been simplified to a pointlike four-fermion one, while still maintaining the symmetry of QCD and the same quantum numbers of the one-gluon exchange. The free-energy density $\cal F$ of this system is a functional of the gap functions and can be obtained from the partition function,
\begin{eqnarray}
\nonumber{\cal Z}&=&{\cal N}e^{-\beta V({\cal F}+(B^2/8\pi))}\\
&=&\int{\cal D}{\bar\psi}{\cal D}\psi\exp\left[\int d^4x({\cal L}-\frac{B^2}{8\pi})\right],\label{functional}
\end{eqnarray}
where $\cal N$ is the normalization constant, $V$ is the 3-volume of the system, and $\beta$ is the inverse absolute temperature. %Notice that $B$ is different from the external magnetic field $H$, since the induced field $B$ includes the medium magnetization.
The Lagrangian density reads
\begin{equation}
{\cal L}=\bar\psi(i\slashed\partial+\tilde{e}\tilde{Q}\slashed {\widetilde{A}}+\mu\gamma_0)\psi-G(\bar\psi\Gamma^a_\mu\psi)(\bar\psi\Gamma_a^\mu\psi).
\label{lagrangian}
\end{equation}
where $\widetilde{A}_\mu$ is the in-medium electromagnetic field corresponding to the symmetry $U(1)_{\tilde Q}$ with coupling constant ${\tilde e}=e\cos\theta$, and $\psi$ is the quark spinor with implicit color and flavor indices. The shortcoming of this model is that the results will not converge in the ultraviolet region because of the lack of asymptotic freedom, which is a significant asset of QCD. Hence, we have to add this feature by hand through introducing a UV momentum cutoff $\Lambda$. Moreover, we have another free parameter in this model, which is the four-fermion coupling constant $G$. These two parameters ($\Lambda$ and $G$) will be fixed to reproduce the CFL gap $\Delta_0$ at a reference chemical potential for $B=0$. In the following, we take the chemical potential and the momentum cutoff to be $500MeV$ and $1GeV$, respectively. Accordingly, the coupling constant will be set to yield the CFL gap $\Delta_0=25MeV$.

It is a little more involved than usual to obtain the particle-particle channels for the Lagrangian density (\ref{lagrangian}) via Fierz transformations. As it is known, the standard Lorentz-covariant basis of $4\times 4$ matrices is created from the combinations of scalar, vector, tensor, axial-vector and pseudo-scalar structures formed from the Dirac $\gamma$-matrices, i.e. $\{1,\gamma^\mu, \sigma^{\mu\nu}, \gamma^\mu\gamma_5, i\gamma_5\}$. Now in a CS, Lorentz symmetry is broken down by the presence of the Fermi sea to a mere rotational symmetry. Furthermore, the rotational symmetry is partially broken by the in-medium uniform magnetic field to rotations in the plane perpendicular to the field direction, which we assume points in the $\hat z$-direction. Therefore, we need to work with the basis matrices given by $\{1, \gamma_0, \gamma^a, \gamma^3, \sigma^{a0}, \sigma^{30}, \sigma^{ab}, \sigma^{3a}, \gamma_0\gamma_5, \gamma^a\gamma_5, \gamma^3\gamma_5, i\gamma_5\}$ with $a,b=1,2$. This procedure will yield to (see details in appendix A)
\begin{equation}
{\cal L}_{\rm int}=G^\prime(\bar\psi P_\eta{\bar\psi}^t)(\psi^t {\bar P}_\eta\psi)+G^{\prime\prime}(\bar\psi M_\rho{\bar\psi}^t)(\psi^t {\bar M}_\rho\psi),
\label{pairingeq}
\end{equation}
with
\begin{equation}
(P_\eta)^{\alpha\beta}_{ij}=i{\cal C}\gamma_5\epsilon^{\alpha\beta\eta}\epsilon_{ij\eta},\qquad \eta=1,2,3,
\end{equation}
and
\begin{equation}\label{Mats}
(M_r)^{\alpha\beta}_{ij}={\cal C}\gamma^5 \sigma_{ab}\epsilon^{\alpha\beta r}(\delta_{is}\delta_{jt}+\delta_{it}\delta_{js}),
\end{equation}
where in (\ref{Mats}) the indices $r,s,t$ take values from $1,2,3$, keeping a cyclic order. Notice that due to the symmetry breaking in the presence of an external magnetic field, the coefficients for both channel will be slightly different. However, our numerical calculations show that the results for all the gaps insignificantly depend on the difference between these two coefficients. Thus, we will consider $G^\prime=G^{\prime\prime}=G_0$ in the following procedures. Next, for each channel we introduce a complex scalar field $\phi_\eta$ and $\chi_\rho$ respectively, and write the interaction Lagrangian as
\begin{eqnarray}
\nonumber
{\cal L}_{\rm int}=&-&\frac{\phi^*_\eta\phi_\eta+\chi^*_\rho\chi_\rho}{G_0}\\
\nonumber&+&\frac{1}{2}(\bar\psi P_\eta{\bar\psi}^T)\phi_\eta+\frac{1}{2}\phi^*_\eta(\psi^T {\bar P}_\eta\psi)\\
&+&\frac{1}{2}(\bar\psi M_\rho{\bar\psi}^T)\chi_\rho+\frac{1}{2}\chi^*_\rho(\psi^T {\bar M}_\rho\psi).
\end{eqnarray}

Then, using a Hubbard-Stratonovich transformation, introducing for the complex scalar fields $\phi_\eta$ and $\chi_\rho$ the corresponding expectation values $\Delta_\eta$ and $({\Delta_M})_\rho$, and taking the mean-field approximation (i.e. neglecting the field fluctuation terms with order higher than two), we obtain that the functional integral (\ref{functional}) becomes quadratic in the quark fields. For the convenience of calculations, we make a rotation so (block) diagonalizing the gap matrix
\begin{widetext}
\begin{equation}
\left(
\begin{array}{ccccccccc}
0 & {\hat\Delta} & {\hat\Delta}_B+{\hat \Delta_M} & 0 & 0 & 0 & 0 & 0 & 0\\
{\hat\Delta} & 0 & {\hat\Delta}_B+{\hat \Delta_M} & 0 & 0 & 0 & 0 & 0 & 0\\
{\hat\Delta}_B-{\hat \Delta_M} & {\hat\Delta}_B-{\hat \Delta_M} & 0 & 0 & 0 & 0 & 0 & 0 & 0\\
0 & 0 & 0 & 0 & -{\hat\Delta} & 0 & 0 & 0 & 0\\
0 & 0 & 0 & -{\hat\Delta} & 0 & 0 & 0 & 0 & 0\\
0 & 0 & 0 & 0 & 0 & 0 & 0 & -{\hat\Delta}_B+{\hat \Delta_M} & 0\\
0 & 0 & 0 & 0 & 0 & 0 & 0 & 0 & -{\hat\Delta}_B+{\hat \Delta_M}\\
0 & 0 & 0 & 0 & 0 & -{\hat\Delta}_B-{\hat \Delta_M} & 0 & 0 & 0\\
0 & 0 & 0 & 0 & 0 & 0 & -{\hat\Delta}_B-{\hat \Delta_M} & 0 & 0
\end{array}\right).\;
\label{gapmatrix}
\end{equation}
\end{widetext}
which is in the basis ($s_b, d_g, u_r, s_g, d_b, u_b, u_g, s_r, d_r$). Introducing Nambu-Gorkov spinors
\begin{equation}
\Psi_{(\tilde Q)}=\left(
\begin{array}{c}
\psi_{(\tilde Q)}\\
{\psi}_{{\cal C}(-\tilde Q)}
\end{array}\right),
\label{NGspinor}
\end{equation}
where $\psi_{(\tilde Q)}=\Omega_{(\tilde Q)}\psi$, with charge projectors for the field representation $\psi^t=$($s_b, d_g, u_r, s_g, d_b, u_b, u_g, s_r, d_r$) given by
\begin{subequations}
\begin{eqnarray}
\Omega_{(0)}={\rm diag}(1,1,1,1,1,0,0,0,0),\\
\Omega_{(+)}={\rm diag}(0,0,0,0,0,1,1,0,0),\\
\Omega_{(-)}={\rm diag}(0,0,0,0,0,0,0,1,1).
\end{eqnarray}
\end{subequations}
and satisfying
\begin{equation}
\Omega_{(\eta)}\Omega_{(\eta^\prime)}=\delta_{\eta\eta^\prime}\Omega_{(\eta)},\hspace{0.2cm} \eta,\eta^\prime=0,+,-
\end{equation}
\begin{equation}
\Omega_{(0)}+\Omega_{(+)}+\Omega_{(-)}=1
\end{equation}
the full Lagrangian density can be written as
\begin{equation}\label{Full-Lag}
{\cal L}(x)=\frac{1}{2}{\bar\Psi}_{(\tilde Q)}(x){\cal S}^{-1}_{(\tilde Q)}(x)\Psi_{(\tilde Q)}(x)-\frac{\Delta^2+2\Delta_B^2+2{\Delta_M}^2}{G_0},
\end{equation}
Moreover, based on the ansatz in (\ref{generalansatz}) we already set $\Delta_1=\Delta,$  $\Delta_2=\Delta_3=\Delta_B$ and $({\Delta_M})_1=(\Delta_M)_2=\Delta_M$,  $(\Delta_M)_3=0$. The inverse full propagator reads
\begin{equation}
{\cal S}^{-1}_{(\tilde Q)}(x)=\left(
\begin{array}{cc}
[G^+_{(\tilde Q)0}]^{-1} & \Phi^-_{(\tilde Q)}\\
\Phi^+_{(\tilde Q)} & [G^-_{(\tilde Q)0}]^{-1}
\end{array}\right),
\label{fullprop}
\end{equation}
where ${\tilde Q}=\{0,\pm\}$ is standing for neutral, positively or negatively $\tilde Q$ charged quarks.

In (\ref{fullprop}) we have the inverse propagators
\begin{equation}
[G^{\pm}_{(\tilde Q)0}]^{-1}=[\slashed\Pi_{(\tilde Q)}\pm\mu\gamma_0]\delta^4(x-y),
\end{equation}\label{Inv-Prp}
with
\begin{equation}
\slashed\Pi_{(\tilde Q)}=i\slashed\partial+{\tilde e}{\tilde Q}\slashed A,
\end{equation}\label{Cov-Der}
and off-diagonal elements
\begin{equation}
\Phi^-_{(\tilde Q)}=\gamma_0[\Phi^+_{(\tilde Q)}]^\dagger\gamma_0,
\label{Gap-Func}
\end{equation}
with the gap matrix $\Phi^+_{(\tilde Q)}$, obtained from (\ref{generalansatz}) after a convenient rotation in color and flavor space, and given by
\begin{equation}
\Phi^+_{(\pm)}=\left(
\begin{array}{cc}
\pm{{\hat\Delta}_M}-{\hat\Delta_B} & 0\\
0 & \pm{{\hat\Delta}_M}-{\hat\Delta_B}
\end{array}\right),
\label{chargedmatrix}
\end{equation}
for the charged sector, and by
\begin{equation}
\Phi^{(1)+}_{(0)}=\left(
\begin{array}{cc}
-{\hat\Delta} & 0\\
0 & -{\hat\Delta}
\end{array}\right),
\end{equation}
and
\begin{equation}
\Phi^{(2)+}_{(0)}=\left(
\begin{array}{ccc}
0 & {{\hat\Delta}_M}+{\hat\Delta} & {{\hat\Delta}_M}+{\hat\Delta}_B\\
-{{\hat\Delta}_M}+{\hat\Delta} & 0 & {{\hat\Delta}_M}+{\hat\Delta}_B\\
-{{\hat\Delta}_M}+{\hat\Delta}_B & -{{\hat\Delta}_M}+{\hat\Delta}_B & 0
\end{array}\right).
\label{neutralmatrix}
\end{equation}
for the neutral sector.
In the following, we will deal with each different block separately. The elements of the matrices (\ref{chargedmatrix})-(\ref{neutralmatrix}) are in color-flavor space.

The transformation of the full quark propagator formed by the $\tilde Q$-charged quarks to momentum space can be obtained by using a method originally developed for charged fermions in \cite{Ritus} and later extended to charged vector fields in \cite{Ritus2}. In this approach, the diagonalization in momentum space of the Green functions of charged fermions in the presence of an external electromagnetic field is carried out with the help of the eigenfunction matrices $\textbf{E}^{l(\pm)}_p(x)$, which are the wave functions of the asymptotic states of positive $(+)$, and negative $(-)$ charged fermions in a uniform electromagnetic field and play the role in a magnetized medium of the usual plane-wave Fourier functions at zero field. The transformation functions $\textbf{E}^{l(\pm)}_p(x)$ for charged quark fields are calculated as the solutions of the eigenvalue equation
\begin{equation}
(\Pi_{(\pm)}\cdot\gamma)\textbf{E}^{l(\pm)}_p(x)=\textbf{E}^{l(\pm)}_p(x)(\gamma\cdot{\bar p}_{(\pm)}),
\label{eigenvalue}
\end{equation}
with ${\bar p}_{(\pm)}$ given by
\begin{equation}
{\bar p}_{(\pm)}=(p_0, 0, \pm\sqrt{2{\tilde e}{\tilde B}{l}},p_3),
\label{pbar}
\end{equation}
where $l=0,1,2,...,$ denotes the Landau level numbers of the particle with rotated charge $\widetilde{q}$ in the presence of the magnetic field
\begin{equation}
l=n-sgn(\widetilde{q})\frac{\sigma}{2}+\frac{1}{2},
\end{equation}
with $n=0,1,2,...,$ and $\sigma=\pm 1$ are the spin projections.

The $\textbf{E}^{l(\pm)}_p(x)$ functions are then given by
\begin{subequations}\label{Efunction}
\begin{equation}
\textbf{E}^{l(+)}_p(x)=E^{l(+)}_{p}\Delta(+)+E^{l-1(+)}_{p}\Delta(-),
\end{equation}
\begin{equation}
\textbf{E}^{l(-)}_p(x)=E^{l-1(-)}_{p}\Delta(+)+E^{l(-)}_{p}\Delta(-),
\end{equation}
\end{subequations}
with eigenfunctions
\begin{equation}\label{RF}
E^{l
(\pm)}_{p}(x)={\cal N}_{l}e^{-i(p_0x^0+p_2x^2+p_3x^3)}D_{l}(\rho_{(\pm)}),
\end{equation}
where $D_{l}(\rho_{(\pm)})$ are the parabolic cylinder functions with argument $\rho$ defined by
\begin{equation}
\rho_{(\pm)}=\sqrt{2{\tilde e}{\tilde B}}(x_1\pm p_2/{\tilde e}{\tilde B}),
\end{equation}
The normalization constant ${\cal N}_{l}$ in (\ref{RF}) is
\begin{equation}
{\cal N}_{l}=(4\pi{\tilde e}{\tilde B})^{1/4}/\sqrt{l!},
\end{equation}
and the spin projectors $\Delta(\pm)$ in (\ref{Efunction}) are defined as
\begin{equation}
\Delta(\pm)=\frac{1\pm i\gamma_1\gamma_2}{2}.
\end{equation}
One can show that the $\textbf{E}^{l(\pm)}_p$ functions satisfy the orthogonality condition \cite{Wang}
\begin{equation}\label{orthogonality}
\int d^4x \bar{\textbf{E}}^{l(\pm)}_p(x)\textbf{E}^{l'(\pm)}_p(x)=(2\pi)^4{\hat\delta}^{(4)}(p-p^\prime)\Xi(l),
\end{equation}

with $\bar{\textbf{E}}^{l(\pm)}\equiv\gamma_0(\textbf{E}^{l(\pm)}_p)^{\dagger}\gamma_0$,
\begin{equation}
{\hat\delta}^{(4)}(p-p^\prime)=\delta^{ll^\prime}\delta(p_0-p_0^\prime)\delta(p_2-p_2^\prime)\delta(p_3-p_3^\prime),
\end{equation}
and
\begin{equation}\label{zerolandau}
\Xi(l)=\Delta({sgn}({\tilde e}{\tilde B}))\delta^{l0}+I(1-\delta^{l0}).
\end{equation}

Using the $\textbf{E}^{l(\pm)}_p(x)$ functions, the charged fields $\psi_{(\pm)}$ can be transformed according to
\begin{subequations}
\begin{equation}
\psi_{(\pm)}(x)=\int\hspace{-0.53cm}\sum\frac{d^4p}{(2\pi)^4}\textbf{E}^{l(\pm)}_p(x)\psi_{(\pm)}(p),
\end{equation}
\begin{equation}
\bar{\psi}_{(\pm)}(x)=\int\hspace{-0.53cm}\sum\frac{d^4p}{(2\pi)^4} \bar{\psi}_{(\pm)}(p) \bar {\textbf{E}}^{l(\pm)}_p(x).
\end{equation}
\label{transform}
\end{subequations}
where we have defined $\int\hspace{-0.38cm}\sum\frac{d^4p}{(2\pi)^4}\equiv\sum^\infty_{n=0}\int\frac{dp_0dp_2dp_3}{(2\pi)^4}$.

The relations (\ref{transform}) and (\ref{eigenvalue}), as well as the orthogonality condition (\ref{orthogonality}), enable us to transform the space dependent part of the Lagrangian (\ref{Full-Lag}) into momentum space
\begin{equation}
\int{\cal L}d^4x=\frac{1}{2}\int\hspace{-0.53cm}\sum\frac{d^4pd^4p'}{(2\pi)^4}{\bar\Psi}_{(\tilde Q)}(p){\cal S}^{-1}_{(\tilde Q)}(p,p')\Psi_{(\tilde Q)}(p'),
\end{equation}
where
\begin{equation}
{\cal S}_{(\tilde Q)}^{-1}(p,p')=\left(
\begin{array}{cc}
[G^{l+}_{(\tilde Q)0}(p,p')]^{-1} & \Phi^-_{(\tilde Q)}\widehat{\delta}^{(4)}(p-p')\\
\Phi^+_{(\tilde Q)}\widehat{\delta}^{(4)}(p-p') & [G^{l-}_{(\tilde Q)0}(p,p')]^{-1}
\end{array}\right),
\end{equation}
The bare inverse propagators in momentum space are given by
\begin{equation}\label{full-FP}
[G^{l\pm}_{(\tilde Q)0}(p,p')]^{-1}=(2\pi)^4\widehat{\delta}^{(4)}(p-p')\Xi(l)[\widetilde{G}^{l\pm}_{(\tilde Q)0}({\bar p}_{(\tilde Q)})]^{-1}
\end{equation}
with $[\widetilde{G}^l_{(\tilde Q)0}({\bar p}_{(\tilde Q)})]^{-1}$ formally
given for neutral quarks by
\begin{equation}
[G^\pm_{(0)0}({\bar p}_{(0)})]^{-1}={\slashed{\bar p}}_{(0)}\pm\mu\gamma_0,
\end{equation}
where ${\bar p}_{(0)}=(p_0,p_1,p_2,p_3)$ is the usual 4-momentum of a free particle, and for the charged quarks by
\begin{subequations}
\begin{equation}
[G^\pm_{(+)0}({\bar p}_{(+)})]^{-1}=\slashed{\bar p}_{(+)}\pm\mu\gamma_0,
\end{equation}
\begin{equation}
[G^\pm_{(-)0}({\bar p}_{(-)})]^{-1}=\slashed{\bar p}_{(-)}\pm\mu\gamma_0.
\end{equation}
\label{chargedpropg}
\end{subequations}
with the double sign in ${\bar p}_{(\pm)}$ denoting either positively or negatively charged particle.

From (\ref{full-FP}), we see that because of the presence of the operator $\Xi(l)$, the lowest Landau level (LLL) is  automatically separated from the rest of the levels. In the MCFL phase, this separation gives rise to a spin degeneracy factor for the $l$ Landau levels, $g_l=2-\delta_{l0}$, in the free energy, reflecting the fact that higher LL's ($l>0$) are double degenerated. However, in the present case the situation is different. The magnetic field interaction with the pair magnetic moment breaks the spin degeneracy of the nonzero LLs, a fact that will be reflected in the dispersion of the charged quasiparticle excitations in the background of the condensate and also in the different way that ${\Delta_M}$ enters in the off-diagonal part of the inverse propagator for zero and nonzero LLs. We will discuss more about this point in the Appendix C.

Integrating over the Nambu-Gorkov spinors, we obtain the partition function as
\begin{equation}
{\cal Z}=\prod_{\tilde Q}\left[\text{Det}\frac{{\cal S}^{-1}_{(\tilde Q)}(i\omega_k,{\bf p})}{1/\beta}\right]^{1/2}\text{exp}(-\beta V\frac{\Delta^2+2\Delta_B^2+2{\Delta_M}^2}{G_0}),
\label{partition}
\end{equation}
and hence the system free energy $\Omega=-\frac{1}{\beta}lnZ$ is found to be
\begin{eqnarray}\label{Omega-2}
\nonumber
\Omega=&-&\frac{1}{\beta}\sum_{\tilde Q}\int\hspace{-0.53cm}\sum\frac{d^4 p}{(2\pi)^4}\frac{1}{2}\text{Tr}\ln[\beta{\cal S}^{-1}_{(\tilde Q)}(i\omega_k, {\bf p})]\\
&+&\frac{\Delta^2+2\Delta_B^2+2{\Delta_M}^2}{G_0}.
\end{eqnarray}
where $\omega_k=(2k+1)\pi/\beta$, $k=0\pm1,\pm2,...$, are the Matsubara frequencies. The Matsubara sum in (\ref{Omega-2}) can be evaluated using the identity
\begin{equation}\label{sum}
\frac{1}{\beta}\sum_k\text{ln}(\frac{\omega_k^2+\varepsilon^2}{1/\beta^2})=|\varepsilon|+\frac{2}{\beta}\text{ln}(1+e^{-\beta|\varepsilon|}).
\end{equation}
In the zero-temperature limit, only the first term from the RHS of (\ref{sum}) survives, leading to the result
\begin{eqnarray}
\nonumber
\Omega=&-&\int_\Lambda\frac{d^3p}{(2\pi)^3}\frac{1}{2}\sum_{j=1}^2|\varepsilon_j|-{\sum_{l=0}^{n_{\widetilde{B}}}} \int_\Lambda \frac{dp_2dp_3}{(2\pi)^3}|\varepsilon^c|\\
&+&\frac{\Delta^2+2\Delta_B^2+2{\Delta_M}^2}{G_0}.
\label{freeenergy}
\end{eqnarray}
where $\Lambda$ is the energy cutoff of the NJL effective theory and $n_{\widetilde{B}}=I[\Lambda^2/2\widetilde{e}\widetilde{B}]$, with $I[...]$ denoting the integer part of the argument. In (\ref{freeenergy}), $\varepsilon_j$ and $\varepsilon^c$ are the energy modes of the neutral and charged quasiparticles respectively, i.e., the values of the energy on which the pole of the determinants laid
\begin{equation}
\text{det}{\cal S}_{(0)}^{-1}[i\varepsilon_j, {\bf p}]=\text{det}{\cal S}_{(\pm)}^{-1}[i\varepsilon^c, {\bf p}]=0.
\label{findzero}
\end{equation}

Notice that we use a color-flavor basis in which the gap matrix (\ref{gapmatrix}) is conveniently block diagonal. Therefore, the full inverse propagator is also block diagonal, and the determinant can be broken into four manageable pieces in (\ref{partition}). Using the identity
\begin{equation}\label{determinante}
\text{det}\left(
\begin{array}{cc}
A & B\\
C & D
\end{array}\right)=\text{det}(A-BD^{-1}C)\text{det}D,
\end{equation}
we obtain the dispersion relations of the different modes by finding the zeros of the determinants in (\ref{findzero}), and counting the corresponding degeneracy $d$.
In this way, we find
\begin{equation}\label{NM-1}
\varepsilon_1=\pm\sqrt{(p\pm\mu)^2+\Delta^2}\hspace{0.6cm}(d=6),
\end{equation}
and
\begin{widetext}
\begin{eqnarray}
\nonumber
\varepsilon_2=&\pm&\left[2(\Delta_B^2+{\Delta_M}^2)+\frac{1}{2}\Delta^2+(p_3^2+p_\perp^2)+\mu^2\right.\\
&\pm&\frac{1}{2}\left.\sqrt{\Delta_a^2\Delta_b^2+16[2{\Delta_M}^2p_\perp^2+\mu^2(p_\perp^2+p_3^2)]\pm8\mu\sqrt{\Delta_b^2(\Delta_a^2p_3^2+\Delta^2p_\perp^2)}}\right]^{1/2}(d=2),
\label{dispersionneutral}
\end{eqnarray}
\end{widetext}
for neutral quarks. In (\ref{dispersionneutral}), we defined $\Delta_a^2=\Delta^2+8{\Delta_M}^2$ and $\Delta_b^2=\Delta^2+8\Delta_B^2$.
The dispersion relations for charged quarks in higher LL's $(l> 0)$ are
\begin{widetext}
\begin{equation}
\varepsilon^c=\pm\sqrt{2{\tilde e}{\tilde B}l+{\Delta_M}^2+\Delta_B^2+\mu^2+p_3^2\pm2\sqrt{2{\tilde e}{\tilde B}l({\Delta_M}^2+\mu^2)+(\mu p_3\pm {\Delta_M}\Delta_B)^2}}\hspace{0.6cm}(d=4),
\label{dispersioncharged}
\end{equation}
\end{widetext}
with $p_\perp^2=p_1^2+p_2^2$. Taking into account the degeneracy $d$ of each mode and the products of all sign combinations, we have that the total number of modes in (\ref{NM-1})-(\ref{dispersioncharged}) is $72$, in agreement with the total number of degrees of freedom of the fields defined in $color\times flavor\times Dirac\times Nambu-Gorkov$ space. The double sign in front of ${\Delta_M}$ in (\ref{dispersioncharged}) reflects the breaking of the spin degeneracy for the higher LL modes of the charged quasi-particles due to the presence of the magnetic-moment condensate ${\Delta_M}$.

Clearly, the dispersion relation for the LLL  cannot be found as the limit (l=0) of (\ref{dispersioncharged}), since the LLL does not have spin degeneracy. As in the case of massless QED with a dynamically generated anomalous magnetic moment \cite{magmoment}, the dispersion relation for the charged fermions in the LLL has to be found independently in a reduced space with only one spin projection. This is done in Appendix C where we obtain the LLL dispersion mode
 \begin{equation}\label{E-LLL}
\varepsilon^c_0=\pm\sqrt{(p_3-\mu)^2+(\Delta_B-{\Delta_M})^2}.
\end{equation}

It can be readily checked that the dispersion relations (\ref{NM-1})-(\ref{dispersioncharged}) recover the form of those previously found in the MCFL phase if one puts by hand ${\Delta_M}=0$. The explicit breaking of the rotational symmetry by the magnetic field is manifested in the separation of the parallel and perpendicular components of momenta in the modes (\ref{dispersionneutral}) and (\ref{dispersioncharged}).

The free energy can be explicitly found by substituting the dispersion relations (\ref{NM-1})-(\ref{E-LLL}) in  (\ref{freeenergy}). A stable phase must minimize the free energy with respect to the variation of the three gap parameters, $\Delta$, $\Delta_B$ and ${\Delta_M}$. This gives rise to the gap equations
\begin{equation}
\frac{\partial\Omega}{\partial\Delta}=\frac{\partial\Omega}{\partial\Delta_B}=\frac{\partial\Omega}{\partial {\Delta_M}}=0.
\label{gapequations}
\end{equation}
These gap equations are quite complicated, even in the strong-magnetic-field limit where only the contribution from the LLL is important, and can only be solved numerically.

\section{Numerical Solutions and Discussion}

As previously analyzed, the integral and sum in (\ref{freeenergy}), and consequently in the gap equations (\ref{gapequations}), must be defined up to a cutoff $\Lambda$. However, as pointed out in \cite{NS}, in order to avoid unphysical discontinuities in many thermodynamical quantities, it is useful to introduce a smooth cutoff function $h_\Lambda$ that should approach $1$ at low energies and $0$ at large ones. In this paper, we will follow the same choice as in \cite{NS}
\begin{equation}
h_\Lambda=\text{exp}(-\xi^2/\Lambda^2),
\end{equation}
where for neutral quarks $\xi=p$ and for charged quarks $\xi=\sqrt{p_3^2+2{\tilde e}{\tilde B}l}$. This Gaussian-like cutoff implements the converge in the momentum integral and sum over Landau levels and will not leads to unphysical discontinuities.
\begin{figure}
\includegraphics[height=3.4in, width=3.8in]{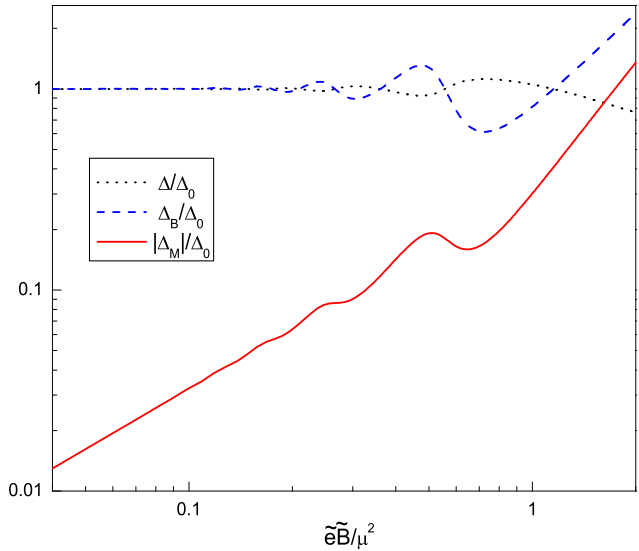}% Here is how to import EPS art
\caption{\label{fig:wide}Gap parameters as a function of ${\tilde e}{\tilde B}/\mu^2$ for $\mu=500MeV$. The corresponding magnetic field for ${\tilde e}{\tilde B}/\mu^2=1$ is $4\times 10^{19}G$. For small magnetic field below those shown in the figure, $\Delta_M/\Delta_0$ behaves as $0.32{\tilde e}{\tilde B}/\mu^2$.}
\label{gaps}
\end{figure}

In Fig.\ref{gaps}, we plot the gaps as functions of a dimensionless parameter ${\tilde e}{\tilde B}/\mu^2$. For small magnetic field, $\Delta$ and $\Delta_B$ are close to each other and approach the CFL gap $\Delta_0=25MeV$. As the magnetic field increases, $\Delta$ and $\Delta_B$  display oscillatory behaviors with respect to ${\tilde e}{\tilde B}/\mu^2$ as long as ${\tilde e}{\tilde B}<\mu^2$. These oscillations are known as the de Hass-van Alphen phenomenon and appear in different charged fermion systems under magnetic fields (see for instance \cite{NS, FW, Klimenko}). As originally explained by Landau \cite{Lev}, these oscillations reveal the quantum nature of the interaction of the charged particles with the magnetic field (what is now called the Landau quantization phenomenon), and are produced by the change in the density of states when passing from one Landau level to another. The oscillations cease when the first Landau level exceeds the Fermi surface. For ultra-strong fields, when only the LLL contributes to the gap equation, $\Delta_B$ is much larger than $\Delta$, which was first found by analytical calculation in \cite{mCFL}. The reason for this phenomenon is that the field increases the density of states of the charged quarks and thus the pairing of charged particles will be reinforced by the penetrating magnetic field.

It is apparent from the graphical representation of $\Delta_M$ in Fig.{\ref{gaps}}, that its value remains relatively small up to magnetic-field values of the order of $\mu^2$. In the field region between $10^{18}-10^{19}G$, the magnitude of $\Delta_M$ grows from a few tenths of Mev to tens of Mev. It becomes comparable to the MCFL gap $\Delta_{B}$ when the field is strong enough to put all the quarks in the LLL, shown in the final segment of the plots in the figure. The de Hass-van Alphen oscillations are much smaller for $\Delta_M$ than for the regular gaps. These features indicate that the main contribution to this gap should come from pairs whose charged quarks are at the LLL. From a physical point of view, this can be understood taking into account that the contribution from higher LL's ($l>0$) should be negligible because the magnetic moment of a pair where the positive quark has spin up and the negative quark has spin down cancels out with that of a pair where the positive quark has spin down and the negative quark has spin up. The only contribution from higher LLs can come when the number of particles is odd, so there are energy states occupied by a single particle, but that is a very small part. The cancelation does not occur, however, between the pairs of quarks in the LLL because they can only be formed by positive quarks with spin up and negative quarks with spin down. At low fields, the number of quarks in the LLL is scarce, while for fields of order ${\tilde e}{\tilde B}\geq \mu^2$, all the particles are constrained to the LLL, hence the variation of $\Delta_M$ from lower values at weak field, to higher values at sufficiently strong fields.

Another important consequence of the new gap is the increment in the magnitude of $\Delta_{B}$ for any given value of the magnetic field in the strong field region, as compared to its own value found at the same field but ignoring the existence of $\Delta_M$. This effect, combined with the increase of $\Delta_M$ at strong fields, will make the CFL phase in the presence of a strong magnetic field more stable than the regular CFL, a fact that could favor the realization of an MCFL core in magnetars.

Our calculations show that at zero magnetic field $\Delta_M$ is zero, so the system does not behave as a ferromagnet. This result is consistent with the rotational invariance of the CFL phase because although the Cooper pairs formed by charged quarks have nonzero magnetic moment even in the absence of a magnetic field, if these moments were spontaneously aligned in some direction, they would break the rotational symmetry of the CFL phase. Accordingly, the expectation value of the magnetic moment at zero field (in the CFL phase) must vanish:
\begin{eqnarray}\label{Vacuum-inv}
\nonumber
\langle0|\Delta_M|0\rangle_{CFL}&=&\langle0|R^{-1}R\Delta_MR^{-1}R|0\rangle_{CFL}\\
&=&\langle0|R\Delta_MR^{-1}|0\rangle_{CFL}=0.\quad
\end{eqnarray}
However, the existence of expectation value of $\Delta_M$ in the presence of a magnetic field is unavoidable. We found that there is no solution of the gap equations with $B\neq 0, \Delta \neq 0, \Delta_B \neq 0$, and $\Delta_M=0$.

%Our new finding is the different from zero average magnetic moment of Cooper pairs formed by charged quarks in the presence of a magnetic field. We want to underline that the magnetic moment of the Cooper pairs of charged quarks should be nonzero even at zero magnetic field, since the paired quarks have opposite charges and opposite spins. But, we should not expect a spontaneous alignment of the magnetic moments, with the consequent induction of a ferromagnetic ground state. The expectation value of the magnetic moment should be zero at $H=0$, as it was found in our numerical calculations (see Fig. 1). This finding is supported by the following symmetry analysis. Recall that the ground state $|0>$ of the CFL phase at zero magnetic field is invariant under rotation $R$, but not the magnetic moment. So the expectation value of the magnetic moment at zero field must vanish:
%\begin{equation}\label{Vacuum-inv}
%<0|T|0>=<0|R^{-1}RTR^{-1}R|0>=<0|RTR^{-1}|0>=0.
%\end{equation}

%Now, once the magnetic field is switched on, the situation is different. As pointed out before, the external magnetic field breaks the rotation invariance of the ground state, so relaxing the condition that gives rise to (\ref{Vacuum-inv}), and then the magnetic moment get a different from zero expectation value. From our numerical calculations we find that a solution with $B\neq 0, \Delta \neq 0, \Delta_H \neq 0$, and $T=0$ is forbidden.

\begin{figure}
\includegraphics[width=0.52\textwidth]{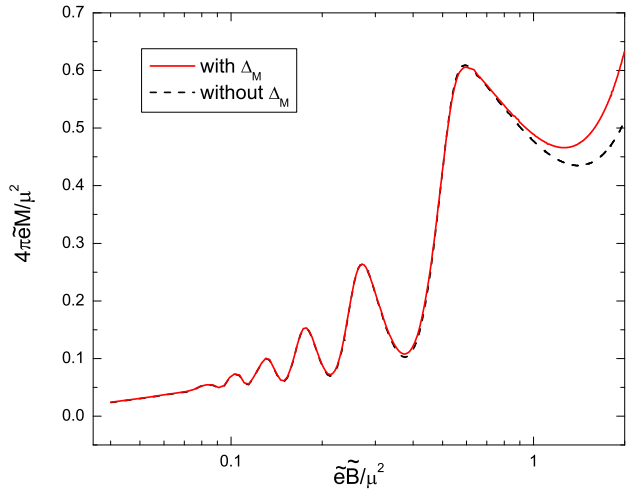}% Here is how to import EPS art
\caption{The ratio $4\pi{\tilde e}M/\mu^2$ as a function of ${\tilde e}{\tilde B}/\mu^2$ with $M$ the magnetization for $\mu=500MeV$ and $\Delta_M\neq 0$ in the solid line; and with $\Delta_M=0$ in the dashed line.}
\label{magnetizationwitht}
\end{figure}

The difference found between the $\Delta's$ and $\Delta_M$ gaps is also in agreement with the fact that the extra order parameter $\Delta_M$ is a consequence of the magnetic field (i.e. $\Delta_M=0$ at $B=0$). In contrast, the relevant scale for the generation of the $\Delta's$ gaps is the energy at the Fermi surface, i.e. the chemical potential. It is logical that once the magnitude of the magnetic field is comparable to the chemical potential, the induced magnetic moment condensate becomes as large as the gap. Another important consequence of the generation of the magnetic moment condensate is that its presence strengthens the gap $\Delta_B$ in the sufficiently strong-magnetic-field region, as can be checked by comparing our results in Fig. 1 with those of Ref. \cite{NS}.

Finally, the magnetization $M=-\partial{\cal F}/\partial B$, which describes the average magnetic moment per unit volume of the superconducting medium at the stationary point, is depicted in Fig. \ref{magnetizationwitht} as a function of the magnetic field. Comparing the two cases, $\Delta_M=0$ (dashed line in Fig. \ref{magnetizationwitht}) and $\Delta_M\neq 0$ (solid line in Fig. \ref{magnetizationwitht}), it is evident that at strong fields the magnetization is reinforced by the magnetic moment of the Cooper pairs. It increases about $10 \%$-$20 \%$ for fields in the range ${\mu}^{2} \lesssim eB \lesssim 2{\mu}^{2}$, as shown in the shift between the two curves in the figure. The magnetization also exhibits, like the gaps, an oscillatory behavior. This larger magnetization at strong fields will be reflected in the equations of state of the magnetized system through the transverse pressure \cite{EoS-H}.\\

\section{Dispersion relations}

\begin{figure}
\includegraphics[width=0.48\textwidth]{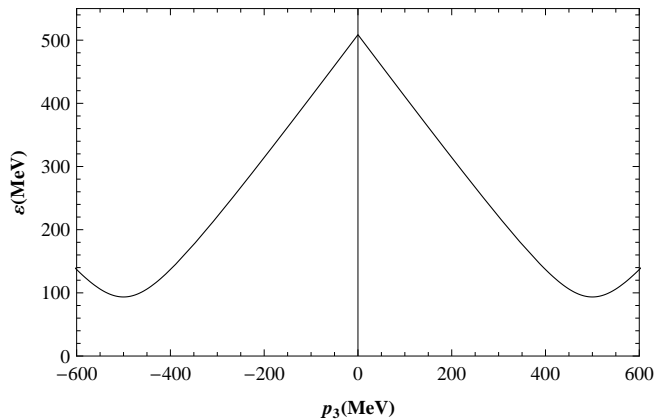}% Here is how to import EPS art
\caption{\label{fig:wide}Dispersion relation for charged quarks at ${\tilde e}{\tilde B}/\mu^2=2$. At that field value the quasiparticles are in the LLL and show a BCS spectrum.}
\label{drchargestrong}
\end{figure}

Let us discuss the effect of the magnetic field on the spectrum of the charged quasiparticles.
In Fig. \ref{drchargestrong}, we plot the dispersion relation of a quasiparticle in the LLL (positive energy in Eq. (\ref{E-LLL})). At the field value considered in the figure, ${\tilde e}{\tilde B}/\mu^2\simeq 2$, the LLL is the only level that fits in the Fermi sphere. As can be seen there, the spectrum of the LLL corresponds to that of the BCS phase with a minimum at $p_3=\mu$.

For higher LL's we need to use the dispersions given in Eq. (\ref{dispersioncharged}). There, the overall double sign denotes particle/hole around the Fermi surface, the double sign in front of the inner square root is for particle/antiparticle, and the double sign in front of $\Delta_M$ corresponds to the two possible projection of the spin of the quasiparticle in the magnetic field. In the case of a quasiparticle, with magnetic moment opposite to the field we have
\begin{widetext}
\begin{equation}
\varepsilon^{'c}=\sqrt{2{\tilde e}{\tilde B}l+\Delta_M^2+\Delta_B^2+\mu^2+p_3^2-2\sqrt{2{\tilde e}{\tilde B}l(\Delta_M^2+\mu^2)+(\mu p_3+\Delta_M\Delta_B)^2}}\quad l\geq 1.
\label{dispersioncharged-2}
\end{equation}
\end{widetext}

\begin{figure}
\includegraphics[width=0.48\textwidth]{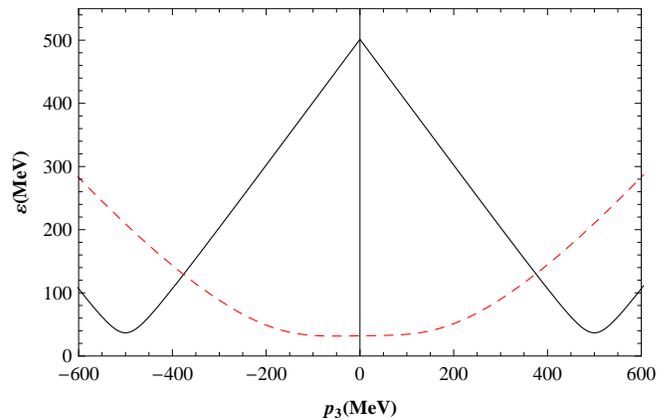}% Here is how to import EPS art
\caption{\label{fig:wide}Dispersion relations for charged quarks at ${\tilde e}{\tilde B}/\mu^2=0.5$. Different lines correspond to the available LL's $l$ at that field value, with black-solid line for $l=0$, and red-dashed line for $l=1$. The upper available LL ($l=1$) has a BEC spectrum.}
\label{drchargesmall}
\end{figure}

\begin{figure}
\includegraphics[width=0.48\textwidth]{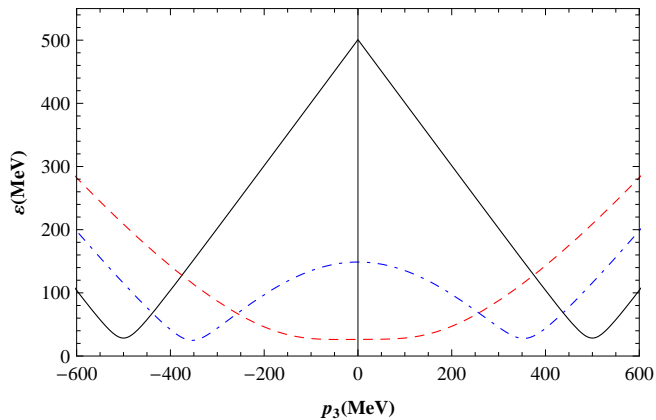}% Here is how to import EPS art
\caption{\label{fig:wide}Dispersion relations for charged quarks at ${\tilde e}{\tilde B}/\mu^2=0.25$. Different lines correspond to different available Landau levels $l$ at that field value. Here, the different spectrum correspond to: black-solid line ($l=0$), blue-dash-dotted line ($l=1$), and red-dashed ($l=2$), respectively. Notice that only the upper available LL has a BEC spectrum.}
\label{drchargesmal2}
\end{figure}

If the value of the magnetic field is such that the Fermi sphere can accommodate $n$ LLs, the dispersion of the last level occupied within the Fermi surface can be either fermionic or bosonic, depending on whether the level is a little below or \emph{exactly at} the Fermi surface.  For example, in the case that the Fermi sphere can exactly accommodate two LL's (i.e. for field values ${\tilde e}{\tilde B}/\mu^2\simeq 0.5$), the level  $l=1$ lies at the Fermi surface. Under this condition, the quasiparticles at l=1 has minimum energy at $p_{3}=0$, a bosonic type of dispersion as shown in Fig. \ref{drchargesmall}. If we decrease the field just a little, $2{\tilde e}{\tilde B}<\mu^{2}$, still there would be only two LLs inside the sphere, but the minimum of the dispersion for $l=1$ would be at a nonzero $p_{3}$, and thus $l=1$ would behave as a fermionic mode. If we continue decreasing the field to ${\tilde e}{\tilde B}/\mu^2\simeq 0.25$ we find the same situation, that is, that the upper LL available in the Fermi sphere (the one with $l=2$ in this case) will have a bosonic spectrum, while the other two have BCS spectra, as can be seen in Fig. \ref{drchargesmal2}. It is easy to understand that the same behavior will be found at any field value. The reason is that once the Fermi sphere is filled with more than one LL, if the upper LL is already at the Fermi surface, then it is only needed an energy equal to the gap to excite a quasiparticle, and consequently the minimum of the dispersion relation takes place at $p_3=0$. The exception occurs when only the LLL fits in the Fermi sphere. In this case, there is no transverse momentum $\overline{p}_\bot = \sqrt{2\widetilde{e}\widetilde{B}l}$ to equate the Fermi sphere radius $\mu$ since $l=0$. Hence, to excite the LLL quasiparticle from the Fermi surface a longitudinal momentum equal to the Fermi sphere radius $p_3=\mu$ is needed, plus the energy gap.

As known, the analytic behavior of the quasiparticle spectrum can determine the nature of the diquark condensate \cite{Leggett, Baym}. It turns out that a magnetic field modifies the quasiparticle dispersions and consequently can change the nature of the diquark pairs. In \cite{BEC-B} the effect of a magnetic field in the dispersions of the quasiparticles in superconductivity was found to be equivalent to producing an effective mass $M_{l}=2eBl$ for the quasiparticles. The relation between that effective mass and the chemical potential determines the fermionic ($\mu > M_{l}$) or bosonic ($\mu < M_{l}$) character of the mode. As discussed in \cite{BEC-B}, a magnetic field can tune the crossover between a BEC and a BCS  regime in a color superconductor by varying the relative numbers of LLs for which the effective chemical potential $\mu_{l}=\mu-M_{l}$ is either positive or negative.

An important outcome of our calculations is the enhancement of the net energy gap of the LLL quasiparticles at strong magnetic fields. As can be seen from (\ref{E-LLL}), the energy gap of the LLL quasiparticles is given by $\Delta_{B}-\Delta_M$. Fig.\ref{gaps} shows that the condensate $\Delta_M$ is negative for all field values, has a small magnitude from low to intermediate fields, and then becomes comparable in magnitude to $\Delta_{B}$ in the region of strong fields. On the other hand, as mentioned above, the gap $\Delta_{B}$, which was found to be enhanced by the magnetic field in the strong field region in the case where $\Delta_M$ was taken zero by hand, becomes even larger in the presence of $\Delta_M$. Therefore, the net energy gap at strong fields is much larger than at zero field. This enhancement effect means that the condensation energy of the pairs of charged quarks is enhanced and this in turn can have implications for the chromomagnetic instability associated to the existence of gapless modes at moderate densities when the mass of the s-quark, $M_s$ cannot be neglected and the color and neutrality conditions are enforced \cite{AKR,Fuku}. The criterion of the phase transition from the CFL to the gapless-CFL phase was found to be \cite{AKR}
\begin{equation}\label{Ms}
\frac{M_s^2}{\mu}\approx 2\Delta_{CFL}.
\end{equation}
If the gap on the RHS of (\ref{Ms}) is enhanced through any effect (in our case it would be due to the presence of a strong magnetic field), the CFL (MCFL if a field is present) phase can bear larger values of $M_s$ without becoming unstable, thus delaying the transition to a gapless phase to even lower baryon densities. Even though the model we considered in our work ignores the effects of the strange quark mass, the arguments used to justify the existence of an extra condensate $\Delta_M$ in the presence of a magnetic field will not change if $M_{s}$ is included.

One can check that the spectrum (\ref{dispersionneutral}), associated to the neutral quasiparticles that couple with the charged ones through the gap equation, has a larger energy gap due to a nonzero $\Delta_M$. Hence,  at strong fields, $\Delta_M$ also produces an enhancement in the energy gap of some of the neutral excitations. We expect that this effect will delay the appearance of gapless excitations for those modes too when the strange mass is included in the analysis.

\section{Concluding remarks and outlook}

In this paper, we extend the study of the MCFL phase of color superconductivity to include the effect of the magnetic moment of the Cooper pairs formed by rotated charged quarks. We found that the external magnetic field aligns the magnetic moments of those pairs producing a net expectation value of the magnetic-moment of the pairs, $\Delta_M/|\widetilde{B}|$, in the field direction. The new order parameter $\Delta_M$ leads to various effects. First, the gap $\Delta_B$ is larger in the strong field region than what it was in \cite{mCFL}, where $\Delta_M$ was not considered. In addition, the net energy gap of the charged quasiparticles in the strong field region is more than twice the one in \cite{mCFL}. Finally, there is no solution that minimizes the MCFL free energy with $\Delta\neq 0$, $\Delta_B\neq 0$ and $\Delta_M=0$. Thus, in the MCFL phase, the magnetic moment condensate has to be considered in equal footing with the gaps $\Delta$ and $\Delta_B$.

The justification for the increase in the number of order parameters in the MCFL phase is easy to understand on the basis of symmetry arguments. The reason is that the magnetic moment condensate $\Delta_M$ does not break any additional symmetry that has not already been broken by the gaps $\Delta$ and $\Delta_B$ and the magnetic field, thus there is no reason why $\Delta_M$ needs to be zero. The situation has some resemblance with the phenomenon of magnetic catalysis in massless QED, where an applied magnetic field explicitly breaks the rotational symmetry and also catalyzes the spontaneous breaking of chiral symmetry through the generation of a chiral condensate. Once these two symmetries are broken there is no reason why any physical parameter whose presence in the Lagrangian would break the exact same symmetries has to be forbidden.  Consequently, two parameters, the mass and the magnetic moment are dynamically generated in the QED case \cite{magmoment}.

Technically, the emergence of a new condensate in the MCFL phase is connected to the modification of the Fierz identities in the presence of a magnetic field. To see this one should first notice that the Lorentz symmetry is broken by the dense medium down to merely a rotational symmetry and, in addition, the magnetic field further breaks this symmetry from $O(3)$ to $O(2)$. When the Fierz transformations are performed in the point-like four-fermion interaction of a NJL theory with these explicitly broken symmetries, one immediately find the opening of new pairing channels, one of which favors the formation of a spin-1 condensate of the form $\Delta_M=\langle \psi^T {\cal{C}} \Sigma^3 \gamma^5 \psi \rangle$, which, as previously discussed, is unavoidable.

We call the reader's attention to the fact that the $\Delta_M$ condensate of the present work is quite different from other spin-1 condensates previously considered in color superconductivity. Usually, spin-1 condensates not only are much smaller than the s-wave ones \cite{Spin-1}, but they do not mix up with s-wave gaps in the quasiparticle spectrum. The existence of quasiparticle dispersions with very small gaps in other phases with spin-1 condensates makes them relevant for the transport properties of the system, but this same characteristic makes these condensates very sensitive to be erased by temperature and, in cases where they break the rotated electromagnetism, also by magnetic fields. In contrast, the spin-1 condensate of our case mix up with the s-wave gaps in the quasiparticle spectrum in such a way that at strong fields it actually increases the effective energy gap of the quasiparticles. Besides, the extra condensate has also the indirect effect of making the gap $\Delta_{B}$ even larger in the strong field region, hence raising the critical temperature required to erase the color superconductivity. This feature makes this phase of potential interest for the planned low-temperature/high-density ion-collision experiments at different worldwide facilities, as pointed out in the Introduction.

In our work we neglected the strange quark mass $M_s$. However, this may not be an accurate assumption for the intermediate densities prevailing in the interior of neutron stars. The response of the CFL superconductor to a nonzero strange quark mass and the constraints imposed by color and electrical neutralities can lead to gapless dispersion relations when the density decreases \cite{AKR}. This in turn has been found to produce chromomagnetic instabilities \cite{Fuku}. We do not see any obvious reason to expect that the enhancement of the energy gap at strong fields found in the present work will not be present too if the strange quark mass is taken into consideration. If the energy gap of some of the quasiparticle dispersions is much larger, obviously the density needed to turn them into gapless modes (and hence produce the chromomagnetic instability) will be much smaller, a fact that could favor the realization of the MCFL phase in the core of a strongly magnetized neutron star, even at intermediate densities. Therefore, an important pending task will be to investigate the MCFL considering a nonzero $M_s$ and imposing the neutrality conditions, while taking into consideration the effect of the Cooper pairs' magnetic moment.

Even if the MCFL phase does not turn out to be the most favored phase in neutron star cores or in the planned experiments, the present investigation is also important from a fundamental point of view, because it has uncovered a mechanism to increase the energy gap that could be also applicable to other color superconducting phases more relevant at intermediate densities. For example, it would be interesting to explore whether a spin-1 condensate similar to the one found here is also favored in the strongly coupled 2SC phase in the presence of a magnetic field. It has been argued that the 2SC phase at strong coupling can be the most energetically favored at moderate densities \cite{scp-2SC}. All the Cooper pairs in this phase are formed by quarks with opposite spins and charges, so they all have nonzero magnetic moments which presumably could be oriented by an external magnetic field. This orientation could lead to a net magnetic moment, likely favoring the formation of a new condensate, as in the MCFL case. If, as in the MCFL situation, the extra condensate in the strongly coupled 2SC theory makes the regular s-wave gap even larger, this will lead to a larger critical temperature, making this phase a potentially interesting candidate for the QCD phases that will be explored by the planned low-temperature/high-density experiments.

From Fig. \ref{magnetizationwitht} we saw that the magnetization is enhanced at strong fields. This feature could serve to determine whether the core of magnetars is made of color superconducting matter or of hadronic matter, whose magnetization is known to be negligible even at large fields \cite{magn-hadr-mat}.

\begin{acknowledgments}
The authors want to express their gratitude to M. Alford, M. Buballa, G. Rupak, D-f, Hou, C. Kouvaris,  J. L. Noronha, K. Rajagopal,  H-c Ren, A. Sanchez, I. A. Shovkovy and H. J. Warringa for helpful discussions and insights. The work of EJF and VI has been supported in
part by DOE Nuclear Theory grant DE-SC0002179.
\end{acknowledgments}

\appendix
\section{The Fierz Identities with both Lorentz Symmetry and Rotational Symmetry Broken}

In this Appendix, we present the details of the Fierz transformation with both Lorentz and rotational symmetry breaking. Fierz identities are connected to reordering of field operators in a contact four-particle interaction. Let us consider an interaction:
\begin{equation}
({\bar\psi_1 A\psi_2})({\bar\psi_3B\psi_4}).
\label{bilinear}
\end{equation}
Here, the indices of the spinors are suppressed. The same interaction can be expressed in a different way as $({\bar\psi_1M\psi_4})({\bar\psi_3N\psi_2})$. How the matrices $A, B$ relate to $M, N$ is the general aim of Fierz transformations.

The $16$ Dirac bilinears in (\ref{bilinear}) are usually classified into distinct classes according to their properties under Lorentz transformation as
\begin{equation}
\{\Gamma^A\}=\{1, \gamma_5, \gamma^\mu, \gamma_5\gamma^\mu, \sigma^{\mu\nu}\},
\label{basis}
\end{equation}
The orthogonality relation for the basis $\{\Gamma^A\}$ are
\begin{equation}
{\text Tr}[\Gamma_A\Gamma^B]=4\delta^B_A,
\label{orthog}
\end{equation}
This relation allows us to expand any complex $4\times 4$ matrix $X$ in terms of the basis (\ref{basis}) as
\begin{equation}
X=X_A\Gamma^A,\quad X_A=\frac{1}{4}{\text Tr}[X\Gamma_A],
\label{expand}
\end{equation}
Combining (\ref{orthog}) and (\ref{expand}), extracting each element of the matrix, we could find a completeness relation as
\begin{equation}
\delta_{mi}\delta_{nj}=\frac{1}{4}(\Gamma_A)_{nm}(\Gamma^A)_{ij}.
\end{equation}
This identity is sufficient to reproduce all possible Fierz identities by appropriately incorporating identity matrices. For general matrices $X, Y$, we have
\begin{eqnarray}
\nonumber
X_{ij}Y_{kl}&=&(X1)_{ij}(1Y)_{kl}=\frac{1}{4}(X\Gamma_C Y)_{il}(\Gamma^C)_{jk}\\
&=&\frac{1}{4^2}{\text Tr}[X\Gamma_CY\Gamma_D](\Gamma^D)_{il}(\Gamma^C)_{jk},
\label{trans}
\end{eqnarray}
In particular, if $X=\Gamma^A$ and $Y=\Gamma^B$, Eq. (\ref{trans}) will lead to the Fierz identities \cite{CC}
\begin{equation}
(\Gamma^A)_{ij}(\Gamma^B)_{kl}=\frac{1}{4^2}\text{Tr}[\Gamma^A\Gamma_C\Gamma^B\Gamma_D](\Gamma^D)_{il}(\Gamma^C)_{jk}.
\label{identity}
\end{equation}
Where, all the lower case letters $i,j,k,l$ run over $0, 1, 2, 3$. The only remaining work is to calculate the expansion coefficients which are straightforwardly obtained as gamma matrix traces. For the particle-particle channel $({\bar\psi}M{\cal C}{\bar\psi}^T)(\psi^T{\cal C}N\psi)$ as in CS, the Fierz identity are related to the coefficients in (\ref{identity}) simply by the sign, which defines the (anti)symmetry of the representation ${\cal C}\Gamma^A$ under transposition (when $A=B$).

For the CFL phase, we should take into account that the Lorentz symmetry is broken by the presence of the dense medium down to mere rotation symmetry. One thus needs to work with the basis of rotation-covariant matrices, $\{1, \gamma_0, \gamma^a, \sigma^{a0}, \sigma^{ab}, \gamma_0\gamma_5, \gamma^a\gamma_5, i\gamma_5\}$, with indices $a$ and $b$ running from one to three. Moreover, for the MCFL phase the rotation symmetry will be further broken by the penetrating magnetic field that is taken along the $\widehat{z}$-direction. Therefore, the expansion basis (\ref{basis}) has to be changed to $\left\{1, \gamma_0, \gamma^a, \gamma^3, \sigma^{a0}, \sigma^{30}, \sigma^{ab}, \sigma^{3a}, \gamma_0\gamma_5, \gamma^a\gamma_5, \gamma^3\gamma_5, \gamma_5\right\}$ with $a, b=1,2$.

Following the same procedure as that in going from (\ref{basis}) to (\ref{identity}), we can obtain the Fierz identities for particle-antiparticle channel as
\begin{widetext}
\begin{equation}
\left(
\begin{array}{c}
(1)_{ij}(1)_{kl}\\
(\gamma_0)_{ij}(\gamma_0)_{kl}\\
(\gamma^a)_{ij}(\gamma_a)_{kl}\\
(\gamma^3)_{ij}(\gamma_3)_{kl}\\
(\sigma^{a0})_{ij}(\sigma_{a0})_{kl}\\
(\sigma^{30})_{ij}(\sigma_{30})_{kl}\\
\frac{1}{2}(\sigma^{ab})_{ij}(\sigma_{ab})_{kl}\\
(\sigma^{3a})_{ij}(\sigma_{3a})_{kl}\\
(\gamma_0\gamma_5)_{ij}(\gamma_0\gamma_5)_{kl}\\
(\gamma^a\gamma_5)_{ij}(\gamma_a\gamma_5)_{kl}\\
(\gamma^3\gamma_5)_{ij}(\gamma^3\gamma_5)_{kl}\\
(i\gamma_5)_{ij}(i\gamma_5)_{kl}
\end{array}\right)=\left(
\begin{array}{rrrrrrrrrrrr}
\frac{1}{4} & \frac{1}{4}  & \frac{1}{4} & \frac{1}{4} & \frac{1}{4} & \frac{1}{4} & \frac{1}{4} & \frac{1}{4} & -\frac{1}{4} & -\frac{1}{4} & -\frac{1}{4}  & -\frac{1}{4} \\
\frac{1}{4} & \frac{1}{4}  & -\frac{1}{4} & -\frac{1}{4} & -\frac{1}{4} & -\frac{1}{4} & \frac{1}{4} & \frac{1}{4} & \frac{1}{4} & -\frac{1}{4} & -\frac{1}{4}  & \frac{1}{4} \\
\frac{1}{2} & -\frac{1}{2}  & 0 & -\frac{1}{2} & 0 & \frac{1}{2} & -\frac{1}{2} & 0 & -\frac{1}{2} & 0 & -\frac{1}{2}  & \frac{1}{2} \\
\frac{1}{4} & -\frac{1}{4}  & -\frac{1}{4} & \frac{1}{4} & \frac{1}{4} & -\frac{1}{4} & \frac{1}{4} & -\frac{1}{4} & -\frac{1}{4} & -\frac{1}{4} & \frac{1}{4}  & \frac{1}{4} \\
\frac{1}{2} & -\frac{1}{2}  & 0 & \frac{1}{2} & 0 & -\frac{1}{2} & -\frac{1}{2} & 0 & \frac{1}{2} & 0 & -\frac{1}{2}  & -\frac{1}{2} \\
\frac{1}{4} & -\frac{1}{4}  & \frac{1}{4} & -\frac{1}{4} & -\frac{1}{4} & \frac{1}{4} & \frac{1}{4} & \frac{1}{4} & \frac{1}{4} & -\frac{1}{4} & \frac{1}{4}  & -\frac{1}{4} \\
\frac{1}{4} & \frac{1}{4}  & \frac{1}{4} & \frac{1}{4} & \frac{1}{4} & \frac{1}{4} & \frac{1}{4} & \frac{1}{4} & -\frac{1}{4} & -\frac{1}{4} & -\frac{1}{4}  & -\frac{1}{4} \\
\frac{1}{2} & \frac{1}{2}  & 0 & -\frac{1}{2} & 0 & -\frac{1}{2} & -\frac{1}{2} & 0 & -\frac{1}{2} & 0 & \frac{1}{2}  & -\frac{1}{2} \\
-\frac{1}{4} & \frac{1}{4}  & -\frac{1}{4} & -\frac{1}{4} & \frac{1}{4} & \frac{1}{4} & -\frac{1}{4} & -\frac{1}{4} & \frac{1}{4} & -\frac{1}{4} & -\frac{1}{4}  & -\frac{1}{4} \\
-\frac{1}{2} & -\frac{1}{2}  & 0 & -\frac{1}{2} & 0 & -\frac{1}{2} & \frac{1}{2} & 0 & -\frac{1}{2} & 0 & -\frac{1}{2}  & -\frac{1}{2} \\
-\frac{1}{4} & -\frac{1}{4}  & -\frac{1}{4} & \frac{1}{4} & -\frac{1}{4} & \frac{1}{4} & -\frac{1}{4} & -\frac{1}{4} & -\frac{1}{4} & -\frac{1}{4} & \frac{1}{4}  & -\frac{1}{4} \\
-\frac{1}{4} & \frac{1}{4}  & \frac{1}{4} & \frac{1}{4} & -\frac{1}{4} & -\frac{1}{4} & -\frac{1}{4} & -\frac{1}{4} & -\frac{1}{4} & -\frac{1}{4} & -\frac{1}{4}  & \frac{1}{4}
\end{array}\right)\left(
\begin{array}{c}
(1)_{il}(1)_{kj}\\
(\gamma_0)_{il}(\gamma_0)_{kj}\\
(\gamma^a)_{il}(\gamma_a)_{kj}\\
(\gamma^3)_{il}(\gamma_3)_{kj}\\
(\sigma^{a0})_{il}(\sigma_{a0})_{kj}\\
(\sigma^{30})_{il}(\sigma_{30})_{kj}\\
\frac{1}{2}(\sigma^{ab})_{il}(\sigma_{ab})_{kj}\\
(\sigma^{3a})_{il}(\sigma_{3a})_{kj}\\
(\gamma_0\gamma_5)_{il}(\gamma_0\gamma_5)_{kj}\\
(\gamma^a\gamma_5)_{il}(\gamma_a\gamma_5)_{kj}\\
(\gamma^3\gamma_5)_{il}(\gamma^3\gamma_5)_{kj}\\
(i\gamma_5)_{il}(i\gamma_5)_{kj}
\end{array}\right),
\end{equation}
\end{widetext}
and that for particle-particle channel as
\begin{widetext}
\begin{equation}
\left(
\begin{array}{c}
(1)_{ij}(1)_{kl}\\
(\gamma_0)_{ij}(\gamma_0)_{kl}\\
(\gamma^a)_{ij}(\gamma_a)_{kl}\\
(\gamma^3)_{ij}(\gamma_3)_{kl}\\
(\sigma^{a0})_{ij}(\sigma_{a0})_{kl}\\
(\sigma^{30})_{ij}(\sigma_{30})_{kl}\\
\frac{1}{2}(\sigma^{ab})_{ij}(\sigma_{ab})_{kl}\\
(\sigma^{3a})_{ij}(\sigma_{3a})_{kl}\\
(\gamma_0\gamma_5)_{ij}(\gamma_0\gamma_5)_{kl}\\
(\gamma^a\gamma_5)_{ij}(\gamma_a\gamma_5)_{kl}\\
(\gamma^3\gamma_5)_{ij}(\gamma^3\gamma_5)_{kl}\\
(i\gamma_5)_{ij}(i\gamma_5)_{kl}
\end{array}\right)=\left(
\begin{array}{rrrrrrrrrrrr}
-\frac{1}{4} & -\frac{1}{4}  & -\frac{1}{4} & -\frac{1}{4} & -\frac{1}{4} & -\frac{1}{4} & -\frac{1}{4} & -\frac{1}{4} & \frac{1}{4} & \frac{1}{4} & \frac{1}{4}  & \frac{1}{4} \\
\frac{1}{4} & \frac{1}{4}  & -\frac{1}{4} & -\frac{1}{4} & -\frac{1}{4} & -\frac{1}{4} & \frac{1}{4} & \frac{1}{4} & \frac{1}{4} & -\frac{1}{4} & -\frac{1}{4}  & \frac{1}{4} \\
\frac{1}{2} & -\frac{1}{2}  & 0 & -\frac{1}{2} & 0 & \frac{1}{2} & -\frac{1}{2} & 0 & -\frac{1}{2} & 0 & -\frac{1}{2}  & \frac{1}{2} \\
\frac{1}{4} & -\frac{1}{4}  & -\frac{1}{4} & \frac{1}{4} & \frac{1}{4} & -\frac{1}{4} & \frac{1}{4} & -\frac{1}{4} & -\frac{1}{4} & -\frac{1}{4} & \frac{1}{4}  & \frac{1}{4} \\
\frac{1}{2} & -\frac{1}{2}  & 0 & \frac{1}{2} & 0 & -\frac{1}{2} & -\frac{1}{2} & 0 & \frac{1}{2} & 0 & -\frac{1}{2}  & -\frac{1}{2} \\
\frac{1}{4} & -\frac{1}{4}  & \frac{1}{4} & -\frac{1}{4} & -\frac{1}{4} & \frac{1}{4} & \frac{1}{4} & \frac{1}{4} & \frac{1}{4} & -\frac{1}{4} & \frac{1}{4}  & -\frac{1}{4} \\
\frac{1}{4} & \frac{1}{4}  & \frac{1}{4} & \frac{1}{4} & \frac{1}{4} & \frac{1}{4} & \frac{1}{4} & \frac{1}{4} & -\frac{1}{4} & -\frac{1}{4} & -\frac{1}{4}  & -\frac{1}{4} \\
\frac{1}{2} & \frac{1}{2}  & 0 & -\frac{1}{2} & 0 & -\frac{1}{2} & -\frac{1}{2} & 0 & -\frac{1}{2} & 0 & \frac{1}{2}  & -\frac{1}{2} \\
\frac{1}{4} & -\frac{1}{4}  & \frac{1}{4} & \frac{1}{4} & -\frac{1}{4} & -\frac{1}{4} & \frac{1}{4} & \frac{1}{4} & -\frac{1}{4} & \frac{1}{4} & \frac{1}{4}  & \frac{1}{4} \\
\frac{1}{2} & \frac{1}{2}  & 0 & \frac{1}{2} & 0 & \frac{1}{2} & -\frac{1}{2} & 0 & \frac{1}{2} & 0 & \frac{1}{2}  & \frac{1}{2} \\
\frac{1}{4} & \frac{1}{4}  & \frac{1}{4} & -\frac{1}{4} & \frac{1}{4} & -\frac{1}{4} & \frac{1}{4} & \frac{1}{4} & \frac{1}{4} & \frac{1}{4} & -\frac{1}{4}  & \frac{1}{4} \\
\frac{1}{4} & -\frac{1}{4}  & -\frac{1}{4} & -\frac{1}{4} & \frac{1}{4} & \frac{1}{4} & \frac{1}{4} & \frac{1}{4} & \frac{1}{4} & \frac{1}{4} & \frac{1}{4}  & -\frac{1}{4}
\end{array}\right)\left(
\begin{array}{c}
({\cal C})_{il}({\cal C})_{kj}\\
(\gamma_0{\cal C})_{il}({\cal C}\gamma_0)_{kj}\\
(\gamma^a{\cal C})_{il}({\cal C}\gamma_a)_{kj}\\
(\gamma^3{\cal C})_{il}({\cal C}\gamma_3)_{kj}\\
(\sigma^{a0}{\cal C})_{il}({\cal C}\sigma_{a0})_{kj}\\
(\sigma^{30}{\cal C})_{il}({\cal C}\sigma_{30})_{kj}\\
\frac{1}{2}(\sigma^{ab}{\cal C})_{il}({\cal C}\sigma_{ab})_{kj}\\
(\sigma^{3a}{\cal C})_{il}({\cal C}\sigma_{3a})_{kj}\\
(\gamma_0\gamma_5{\cal C})_{il}({\cal C}\gamma_0\gamma_5)_{kj}\\
(\gamma^a\gamma_5{\cal C})_{il}({\cal C}\gamma_a\gamma_5)_{kj}\\
(\gamma^3\gamma_5{\cal C})_{il}({\cal C}\gamma^3\gamma_5)_{kj}\\
(i\gamma_5{\cal C})_{il}({\cal C}i\gamma_5)_{kj}
\end{array}\right).
\label{Fierz}
\end{equation}
\end{widetext}

In the following, we will start from the color current interaction to obtain (\ref{pairingeq}). Considering the Lagrangian density (\ref{4fermion-interact})
\begin{eqnarray}\label{A10}
\nonumber{\cal L}=&-&g_E({\bar\psi}\gamma_0\lambda_a\psi)({\bar\psi}\gamma_0\lambda_a\psi)
-g_M^\perp({\bar\psi}\gamma^\perp\lambda_a\psi)({\bar\psi}\gamma_\perp\lambda_a\psi)\\
&-&g_M^3({\bar\psi}\gamma^3\lambda_a\psi)({\bar\psi}\gamma_3\lambda_a\psi),
\end{eqnarray}
which is in agreement with the symmetries if the MCFL phase, and taking into account the Fierz identities for Dirac matrices that can be extracted from (\ref{Fierz}) as
\begin{subequations}
\begin{equation}
(\gamma_0)(\gamma_0)=-\frac{1}{4}\left\{(\sigma^{30}{\cal C})({\cal C}\sigma_{30})-(i\gamma_5{\cal C})(i{\cal C}\gamma_5)+...\right\},
\end{equation}
\begin{equation}
(\gamma^\perp)(\gamma_\perp)=\frac{1}{2}\left\{(\sigma^{30}{\cal C})({\cal C}\sigma_{30})+(i\gamma_5{\cal C})(i{\cal C}\gamma_5)+...\right\},
\end{equation}
\begin{equation}
(\gamma^3)(\gamma_3)=-\frac{1}{4}\left\{(\sigma^{30}{\cal C})({\cal C}\sigma_{30})-(i\gamma_5{\cal C})(i{\cal C}\gamma_5)+...\right\},
\end{equation}
\end{subequations}
where, the Dirac indices have been suppressed. The Fierz identities for the generators of $SU(N)$ can be found in \cite{Buballa}.

For particle-particle channels, we have
\begin{equation}
\left(
\begin{array}{cc}
(1)_{\alpha\beta} & (1)_{\rho\tau}\\
(\lambda_a)_{\alpha\beta} & (\lambda_a)_{\rho\tau}
\end{array}\right)=\left(
\begin{array}{cc}
\frac{1}{2} & \frac{1}{2}\\
\frac{N-1}{N} & -\frac{N+1}{N}
\end{array}\right)\left(
\begin{array}{cc}
(\lambda_S)_{\alpha\rho} & (\lambda_S)_{\tau\beta}\\
(\lambda_A)_{\alpha\rho} & (\lambda_A)_{\tau\beta}
\end{array}\right),
\end{equation}
Here, the Greek letters run from one to three and $\lambda_S, \lambda_A$ are the symmetric and antisymmetric generators of $SU(N)$ respectively. Taking into account everything together, the 4-fermion Lagrangian density (\ref{A10}) becomes
\begin{widetext}
\begin{equation}
{\cal L}=\frac{N_c+1}{8N_c}\left[G^\prime({\bar\psi}\sigma^{30}{\cal C}\lambda_S\lambda_A{\bar\psi}^T)(\psi^T{\cal C}\sigma_{30}\lambda_S\lambda_A\psi)+G^{\prime\prime}({\bar\psi}i\gamma_5{\cal C}\lambda_A\lambda_{A^\prime}{\bar\psi}^T)(\psi^Ti{\cal C}\gamma_5\lambda_A\lambda_{A^\prime}\psi)\right],
\end{equation}
\end{widetext}
where we have defined the new coefficients $G^\prime=\frac{1}{2}g_E-g_M^\perp+\frac{1}{2}g_M^3$ and $G^{\prime\prime}=g_E+2g_M^\perp+g_M^3$. Incorporating the color-flavor ansatz we proposed in Section II, we can obtain the Lagrangian density we used in (\ref{pairingeq}).

\section{Spin-1 Magnetic-Moment Condensate}

Fermion condensates in relativistic theories have to preserve the Lorentz's covariance of the Lagrangian density. For the chiral condensate, for example, the allowed Dirac structures entering in the particle-antiparticle bilinear are the 16 elements of the so-called Dirac ring
\begin{eqnarray}\label{Chiral-cond}
\nonumber<\overline{\psi}_\alpha M_{\alpha \beta} \psi_\beta>,\quad M_{\alpha\beta}=M^SI+M_\mu^V\gamma^\mu+
\\
+M_{\mu\nu}^T\sigma^{\mu\nu}+M_\mu^A\gamma_5\gamma^\mu+M^Pi\gamma_5 \qquad
\end{eqnarray}
Here, the supra-indices in the expansion coefficients, $S, V, T, A$ and $P$, stand for the scalar, vector, tensor, axial vector and pseudoscalar nature of the structure, respectively. An essential property of this decomposition is that each element transforms under a Lorentz transformation into itself. For the superconducting Cooper pairs, the analogous Lorentz decomposition that preserves the relativistic covariance is
\begin{eqnarray}\label{Cooper-cond}
\nonumber<\psi^T_\alpha \Delta_{\alpha \beta} \psi_\beta>,\quad \Delta_{\alpha\beta}=[\Delta^SI+\Delta_\mu^V\gamma^\mu+
\\
+\Delta_{\mu\nu}^T\sigma^{\mu\nu}+\Delta_\mu^A\gamma_5\gamma^\mu+\Delta^Pi\gamma_5](\gamma_5\cal{C}) \qquad
\end{eqnarray}

The magnetic moment condensate corresponds to the tensorial element, $\sigma^{\mu\nu}$, of the Dirac ring in (\ref{Cooper-cond}). Since this condensate is symmetric in Dirac, it has wave functions $|S,M_S\rangle$, with $S$ and $M_S$ representing the pairs' spin and spin projection along the z-axis respectively, that can be given in term of the quarks' $S_z$-spin projection representation, $|m_1,m_2\rangle$ as \cite{Sakurai}
\begin{eqnarray}\label{Wave-functions}
|1,1\rangle=|+,+\rangle, \qquad |1,-1\rangle=|-,-\rangle, \nonumber
\\
|1,0\rangle=\frac{1}{\sqrt{2}}[|+,-\rangle+|-,+\rangle], \qquad \qquad
\end{eqnarray}
The pairs with $|+,+\rangle$ corresponds to quarks with both spins up (+), $|-,-\rangle$ with both spins down (-), and $|\pm,\mp\rangle$ to those with one quark with spin up and the other down, and viceversa.

The spin-1 magnetic-moment condensates in general can be given by
\begin{equation}\label{spin-condensate}
\langle \psi^T {\cal C} \Sigma^{k}\gamma^5 \psi\rangle,\qquad k=1,2,3,
\end{equation}
where $\Sigma^{k}=\frac{1}{2}\varepsilon^{kij}\sigma_{ij}$ is the spin operator.

Introducing the sum of the spin projectors $\Delta(\pm)$ in (\ref{spin-condensate})
\begin{equation}\label{spin-condensate-0}
\langle \psi^T [\Delta(+)+\Delta(-)]{\cal C} \Sigma^{k}\gamma^5[\Delta(+)+\Delta(-)]\psi\rangle,
\end{equation}
and taking into account that $\Delta(\pm)\Delta(\pm)=\Delta(\pm)$, $\Delta(\pm)\Delta(\mp)=0$, we see from (\ref{spin-condensate-0}) that for each spin index $k$, it is obtained
\begin{equation}\label{spin-condensate-1}
\langle \psi^T \Delta(+){\cal C} \Sigma^{1}\gamma^5\Delta(+) \psi\rangle+\langle \psi^T \Delta(-){\cal C} \Sigma^{1}\gamma^5\Delta(-) \psi\rangle,
\end{equation}
\begin{equation}\label{spin-condensate-2}
\langle \psi^T \Delta(+){\cal C} \Sigma^{2}\gamma^5\Delta(+) \psi\rangle+\langle \psi^T \Delta(-){\cal C} \Sigma^{2}\gamma^5\Delta(-) \psi\rangle,
\end{equation}
\begin{equation}\label{spin-condensate-3}
\langle \psi^T \Delta(+){\cal C} \Sigma^{3}\gamma^5\Delta(-) \psi\rangle+\langle \psi^T \Delta(-){\cal C}\Sigma^{3}\gamma^5\Delta(+) \psi\rangle ,
\end{equation}

Because
\begin{equation}\label{Projection}
{\cal C} \Sigma^{1}\gamma^5=\Delta(-)-\Delta(+),\qquad {\cal C} \Sigma^{2}\gamma^5=-iI
\end{equation}
we finally find that
\begin{eqnarray}\label{Projection-1}
|1,1\rangle=-\frac{1}{2}\langle \psi^T {\cal C} \Sigma^{1}\gamma^5 \psi\rangle+\frac{i}{2}\langle \psi^T {\cal C} \Sigma^{2}\gamma^5 \psi\rangle \nonumber
\\
=\langle \psi^T_{(+)}  \psi_{(+)}\rangle, \qquad\qquad\qquad\qquad\qquad
\end{eqnarray}

\begin{eqnarray}\label{Projection-2}
|1,-1\rangle=\frac{1}{2}\langle \psi^T {\cal C} \Sigma^{1}\gamma^5 \psi\rangle+\frac{i}{2}\langle \psi^T {\cal C} \Sigma^{2}\gamma^5 \psi\rangle\qquad\nonumber
\\
=\langle \psi^T_{(-)}  \psi_{(-)}\rangle,\qquad\qquad\qquad\qquad\qquad
\end{eqnarray}
\begin{eqnarray}\label{Projection-3}
|1,0\rangle=\langle \psi^T \Delta(+){\cal C} \Sigma^{3}\gamma^5\Delta(-) \psi\rangle \nonumber
\\
+\langle \psi^T \Delta(-){\cal C}\Sigma^{3}\gamma^5\Delta(+) \psi\rangle \nonumber
\\
=\langle \overline{\psi}_C^{(+)}\gamma^5  \psi^{(-)}\rangle+\langle \overline{\psi}_C^{(-)} \gamma^5 \psi^{(+)}\rangle,
\end{eqnarray}
where $\psi^{(\pm)}$ represents the wave function of the spin-up (+) quarks and spin-down (-) quarks, and $\overline{\psi}_C^{(\pm)}=\psi^T\Delta(\pm)\cal{C}$. We can see that to get the condensate projections $M_S=1, -1$, we need to consider linear combinations of the magnetic-moment condensates along the $x$ and $y$ axes, as one would  expect, since the spin operators in different directions do not commute. The magnetic-moment condensate $\Delta_M$ we are considering in this paper corresponds to the order parameter (\ref{Projection-3}), and consequently to a spin-1 condensate with zero component of the spin $|1,0\rangle$ in the field direction.

\section{The Lowest-Landau-Level Lagrangian}

As we mentioned above, the inverse propagator for the LLL should be different from those for the remaining LL's. In this Appendix, we will justify that statement and obtain the dispersion relations for the quasiparticles in the LLL. Considering now only the contribution of the positively charged quarks, we have
\begin{equation}
{\cal L}=\int d^4x{\bar\Psi}_{(+)}(x){\cal S}^{-1}_{(+)}(x)\Psi_{(+)}(x),
\end{equation}
where the NG spinors and the inverse propagator are given in (\ref{NGspinor}) and (\ref{fullprop}). Expanding this Lagrangian in NG space, we have
\begin{eqnarray}
\nonumber{\cal L}=\int d^4&x&\left[{\bar\psi}_{(+)}(i\slashed\partial+{\tilde e}{\tilde{\slashed A}_{\mu}}+\mu\gamma_0)\psi_{(+)}\right.
\\&+&\left.{\bar\psi}_{{\cal C}(-)}\Phi^+_{(+)}\psi_{(+)}+c.c\right].
\label{nglagrangian}
\end{eqnarray}
Using Ritus' transformation to momentum space for charged fields

\begin{subequations}
\begin{equation}
\psi_{(+)}(x)=\int\hspace{-0.53cm}\sum\frac{d^4p}{(2\pi)^4}\textbf{E}^{l(+)}_p(x)\psi_{(+)}(p),
\end{equation}
\begin{equation}
\bar{\psi}_{(+)}(x)=\int\hspace{-0.53cm}\sum\frac{d^4p}{(2\pi)^4} \bar{\psi}_{(+)}(p)\bar{\textbf{E}}^{l(+)}_p(x).
\end{equation}
\label{transform-0}
\end{subequations}
and taking into account (\ref{eigenvalue}) and (\ref{orthogonality}), we have for the free-propagator part in (\ref{nglagrangian})
\begin{equation}
{\cal L}_F=\int\hspace{-0.53cm}\sum\frac{d^4p}{(2\pi)^4}{\bar\psi}_{(+)}\Xi(l)\left[\gamma_\mu{\bar p}^\mu_{(+)}+\mu\gamma_0\right]\psi_{(+)},
\end{equation}
where $\Xi(l)$ was defined in (\ref{zerolandau}), as the operator that separates the LLL Lagrangian from the rest
\begin{equation}
{\cal L}_F={\cal L}_F^0+\sum_{l=1}^\infty\int\frac{dp_0dp_2dp_3}{(2\pi)^4}\bar{\psi}_{(+)}\left[\gamma_\mu{\bar p}^\mu_{(+)}+\mu\gamma_0\right]\psi_{(+)},
\end{equation}
with the LLL part given by
\begin{equation}\label{B7}
{\cal L}_F^0=\int\frac{dp_0dp_2dp_3}{(2\pi)^4} \bar{\psi}_{(+)}\Delta(+)[\gamma^{||}p_{||}+\mu\gamma_0]\psi_{(+)}.
\end{equation}
The spin projectors $\Delta(\pm)$ can be expressed in terms of Pauli matrices as
\begin{equation}
\Delta(\pm)=\left(
\begin{array}{cc}
\sigma^\pm & \\
 & \sigma^\pm
\end{array}\right),
\end{equation}
where
\begin{equation}
\sigma^\pm=\frac{1}{2}(1+\sigma^3)
\end{equation}
With the chiral projection operators
\begin{equation}
R=\frac{1+\gamma_5}{2}, \hspace{1cm} L=\frac{1-\gamma_5}{2},
\end{equation}
in the chiral representation, $\gamma_5={\rm diag}\{-1,1\}$, we have the following commutation relations
\begin{equation}
[\Delta(\pm),L]=[\Delta(\pm),R]=0.
\end{equation}
In terms of these projectors, we can represent the Dirac spinor as
\begin{equation}
\psi=\psi^{(+)}_R+\psi^{(-)}_R+\psi^{(+)}_L+\psi^{(-)}_L,
\end{equation}
with
\begin{equation}
\psi^{(\pm)}_R=R\Delta(\pm)\psi, \hspace{1cm} \psi^{(\pm)}_L=L\Delta(\pm)\psi.
\end{equation}
and then the Lagrangian density of the LLL (\ref{B7}) can be written as
\begin{eqnarray}
\nonumber
{\cal L}^0_F=& &\int\frac{dp_0dp_2dp_3}{(2\pi)^4}\left\{[{\bar\psi}_{(+)}]^{(+)}_R(\gamma^{||}p_{||}+\mu\gamma_0)[\psi_{(+)}]^{(+)}_R\right.\\&+&\left.[{\bar\psi}_{(+)}]^{(+)}_L(\gamma^{||}p_{||}+\mu\gamma_0)[\psi_{(+)}]^{(+)}_L\right\},
\end{eqnarray}
Here, we have used the relations
\begin{equation}
\gamma^{||}\Delta(\pm)=\Delta({\pm})\gamma^{||}, \hspace{1cm} \gamma^\perp\Delta(\pm)=\Delta(\mp)\gamma^\perp,
\label{zerolagrangian}
\end{equation}
In terms of the explicit components of Dirac spinor $\psi^T_{(+)}=(\psi_1,\psi_2,\psi_3,\psi_4)$, we can rewrite (\ref{zerolagrangian}) as
\begin{eqnarray}
\nonumber{\cal L}_F^0=\int\frac{dp_0dp_2dp_3}{(2\pi)^4}&[&\psi_{1}^*(p_0+\mu+p_3)\psi_1\\
&+&\psi_3^*(p_0+\mu-p_3)\psi_3].
\end{eqnarray}
For the term involving the gaps in (\ref{nglagrangian}), we have
\begin{equation}
{\cal L}_G=\int\hspace{-0.53cm}\sum\frac{d^4p}{(2\pi)^4}{\bar\psi}_{{\cal C}(-)}\Xi(l)\Phi^+_{(+)}\psi_{(+)},
\end{equation}
From where we can isolate the LLL part
\begin{eqnarray}
\nonumber{\cal L}_G^0&=&\int\frac{dp_0dp_2dp_3}{(2\pi)^4}{\bar\psi}_{{\cal C}(-)}\Delta(+)\Phi^+_{(+)}\psi_{(+)}\\
\nonumber&=&\int\frac{dp_0dp_2dp_3}{(2\pi)^4}\left\{[{\bar\psi}_{{\cal C}(-)}]^{(+)}_R\Delta(+)\Phi^+_{(+)}[\psi_{(+)}]^{(+)}_R\right.\\
& &+\left.[{\bar\psi}_{{\cal C}(-)}]^{(+)}_L\Delta(+)\Phi^+_{(+)}[\psi_{(+)}]^{(+)}_L\right\}.
\label{gaplagrangian}
\end{eqnarray}
where we used that $\Delta(+)\Delta(+)=\Delta(+)$.

Notice that the gap matrix $\Phi^+_{(+)}$ can be written as
\begin{equation}\label{B19}
\Phi^+_{(+)}=(-\Delta_B+\Delta_M)\gamma_5\Delta(+)+(-\Delta_B-\Delta_M)\gamma_5\Delta(-),
\end{equation}
Then in (\ref{gaplagrangian}) only the first term in the RHS of (\ref{B19}) can survive since $\Delta(+)\Delta(-)=0$. Thus, we have
\begin{equation}\label{G-Lag}
{\cal L}_G^0=\int\frac{dp_0dp_2dp_3}{(2\pi)^4}[\psi^\prime_4(-\Delta_B+\Delta_M)\psi_3+\psi^\prime_2(-\Delta_B+\Delta_M)\psi_1].
\end{equation}
Here, $\psi^\prime_{2,4}$ are the components of spinor $\psi_{(-)}$. The same procedure can be applied to the charge-conjugate part in (\ref{nglagrangian}). Putting everything together, we have
\begin{equation}\label{LLL-Lag}
{\cal L}^0={\cal L}^0_F+{\cal L}^0_G+c.c=\int\frac{dp_0dp_2dp_3}{(2\pi)^4}{\bar\Psi}_0(p)[{\cal S}_0(p)]^{-1}\Psi_0(p),
\end{equation}
with
\begin{equation}\label{LLL-Spinor}
\Psi_0(p)=\left(
\begin{array}{c}
\psi_1\\
\psi_3\\
\psi_2^{\prime*}\\
\psi_4^{\prime*}
\end{array}\right),
\end{equation}
and
\begin{equation}\label{LLL-Inv-Pro}
[{\cal S}_0(p)]^{-1}=\gamma_0\left(
\begin{array}{cc}
(p_0+\mu)+\sigma_3p_3 & -\Delta_B+\Delta_M\\
-\Delta_B+\Delta_M & (p_0-\mu)-\sigma_3p_3
\end{array}\right).
\end{equation}
Now we could readily obtain the dispersion relations for LLL through ${\rm det}[{\cal S}_0(p)]^{-1}=0$ as
\begin{equation}\label{LLL-Lag}
\varepsilon^c_0=\pm\sqrt{(p_3-\mu)^2+(\Delta_B-\Delta_M)^2}.
\end{equation}
For the negatively charged quarks, the dispersion relations are the same since the sign in (\ref{zerolandau}) will change accordingly.

Note that in the LLL Lagrangian (\ref{LLL-Lag}) the spinor (\ref{LLL-Spinor}) is a 4-component field. This is different from the LLL Lagrangian in QED \cite{magmoment}, where the LLL fermions behaves as the two-component spinors of the (1+1)-D Thirring model \cite{Thirring}. This difference is due to the contribution in the inverse propagator (\ref{LLL-Inv-Pro}) of the Nambu-Gorkov off-diagonal fields corresponding to the charge conjugate of the spin-up negatively charged quarks.

\end{document}